\documentclass[10pt]{article}
\pdfoutput=1
\usepackage{jheppub}
\usepackage{amsmath,amssymb,amsthm,latexsym,bbm,calc,wasysym,mathtools,mathrsfs}
\usepackage[english]{babel}
\usepackage{graphicx,color}
\usepackage[dvipsnames]{xcolor}
\usepackage{tikz-cd}
\usepackage{tikz}
\usetikzlibrary{shapes.geometric}
\usetikzlibrary{arrows,matrix,calc,scopes,decorations.markings}
\usepackage{xspace}
\usepackage{pifont,dsfont}
\usepackage{marvosym}
\usepackage{slashed}
\usepackage{subfigure}
\usepackage{enumitem}
\usepackage{sidecap}

\usepackage[export]{adjustbox}
\mathtoolsset{showonlyrefs=false}

\newcommand{\be}{\begin{equation}}
\newcommand{\ee}{\end{equation}}
\newcommand{\beq}{\begin{equation}}
\newcommand{\eeq}{\end{equation}}
\newcommand{\bes}{\begin{eqnarray}}
\newcommand{\ees}{\end{eqnarray}}

\newcommand\sss{\scriptstyle}

\def\tr{{\textrm{tr} }}

\def\ot{{\otimes}}

\newcommand{\C}{\mathbb{C}}

\newcommand{\q}{\quad}

\newcommand{\cA}{\mathcal{A}}
\newcommand{\cH}{\mathcal{H}}

\newcommand{\la}{\langle}
\newcommand{\ra}{\rangle}
\newcommand{\nn	}{\nonumber}

\newcommand{\hex}{\large\hexagon}
\newcommand{{\rd}}{\rm{d}}
\newcommand{{\su}}{\mathfrak{su}}\newcommand{{\SU}}{\rm{SU}}
\hypersetup{
	colorlinks=true,       
	linkcolor=BlueViolet,          
	citecolor=BrickRed,        
	filecolor=BrickRed,      
	urlcolor=BrickRed           
}

\title{\boldmath Dual loop quantizations of 3d gravity}

\author[a,b]{Clement Delcamp,}
\author[a]{Laurent Freidel,}
\author[c]{Florian Girelli}

\affiliation[a]{Perimeter Institute for Theoretical Physics,\\ 31 Caroline Street North, Waterloo, Ontario  N2L 2Y5, Canada}
\affiliation[b]{Department of Physics $\&$ Astronomy and Guelph-Waterloo Physics Institute \\  University of Waterloo, Waterloo, Ontario N2L 3G1, Canada}
\affiliation[c]{Department of Applied Mathematics \\ University of Waterloo, Waterloo, Ontario N2L 3G1, Canada}

\emailAdd{cdelcamp@perimeterinstitute.ca}
\emailAdd{lfreidel@perimeterinstitute.ca}
\emailAdd{fgirelli@uwaterloo.ca}

\notoc

\abstract{
	The loop quantization of 3d gravity consists in defining the Hilbert space of states satisfying the Gau{\ss} constraint and the flatness constraint. The Gau{\ss} constraint is enforced at the kinematical level by introducing spin networks which form a basis for the Hilbert space of gauge invariant functionals. The flatness constraint is implemented at the dynamical level via the Ponzano-Regge state-sum model.
	We propose in this work a dual loop quantization scheme where the role of the constraints is exchanged. The flatness constraint is imposed first via the introduction of a new basis labeled by group variables, while the Gau{\ss} constraint is implemented dynamically using a projector which is related to the Dijkgraaf-Witten model. We discuss how this alternative quantization program is related to 3d teleparallel gravity.
}

\begin{document} 
	\vspace*{-5em}
	\maketitle
	\flushbottom

\newpage

\section{Introduction}
General relativity  is an example of totally constrained system. Upon quantization, the order in which the constraints are imposed implies a choice of representation for the quantum theory. Changing such order then provides different representations of the same quantum theory. Indeed, each constraint generates a type of symmetry and the physical Hilbert space is defined in terms of representations of such symmetries. It is common to refer to the first constraint to be implemented as the \emph{kinematical} one whereas the last one is the \emph{dynamical} one.
In general, having access to different representations is useful as some physical situations might be better understood in some than others. 

The influence of the order in which the constraints are imposed can be analyzed in great detail in (2+1)d where general relativity is described by a topological field theory, namely $BF$-theory \cite{horowitz1989}. In this context, we have to deal with only two constraints. The first one enforces the flatness of the connection.\footnote{We will work with a zero cosmological constant for simplicity.} The second one, which is often referred to as the \emph{Gau{\ss} constraint}, imposes the torsion-freeness of the connection. It turns out that changing the ordering of the constraints matters when interpreting the theory. Indeed, general relativity can either be seen as a theory about a torsionless connection where the dynamical degrees of freedom are encoded in the curvature, or, as in the teleparallel formulation of gravity \cite{einstein}, as a theory about a flat connection where the dynamical degrees of freedom are encoded in the torsion. In the former case, the kinematical constraint is the Gau{\ss} constraint, while in the latter one, it is the flatness constraint. The Lagrangian associated with these two formulations are related up to a boundary term \cite{Ferraro:2016wht}, however the interpretation of these theories significantly differs: On the one hand gravity is about space-time geometry, and on the other hand, it is still a force. Similarly, in the quantum regime, solving all the constraints in whichever order describes the same theory. Nevertheless, as we will explain, the choice of kinematical constraint specifies the type of representations used to construct the physical Hilbert space.

\bigskip \noindent
The Loop Quantum Gravity (LQG)  quantization program consists in applying Dirac's quantization procedure to general relativity. In (2+1)d, it relies on imposing first the Gau{\ss} constraint at the kinematical level, which implements the local gauge invariance, and then the flatness constraint at the dynamical level, which implements the local translational invariance. Upon quantization, it is natural to choose a representation of the kinematical Hilbert space which makes the Gau{\ss} constraint easy to implement. The result is the so-called \emph{spin network basis} \cite{Rovelli:1995ac} which provides a basis for the Hilbert space of gauge invariant functionals.

The spin network basis which is labeled by SU(2) spin-$j$ representations diagonalizes geometrical operators obtained as the Casimir of the flux operators. Since the spectra of the geometrical operators are discrete, the ground state which is peaked on a totally degenerate geometrical configuration is normalizable. It is the so-called \emph{Ashtekar-Lewandowski} representation \cite{Ashtekar1991, Ashtekar1993, Ashtekar:1994mh}.  Furthermore, the natural local excitations for the spin network basis are pure spin excitations which carries no mass and no momenta while curvature excitations are not easily accessible.

It was suggested recently in \cite{Dupuis:2017otn}, by investigating the discretization of the phase space variables of 3d gravity, that it is possible to impose first the flatness constraint and then the Gau{\ss} constraint at the quantum level. We explore in this manuscript this alternative quantization scheme which we will refer to as \textit{dual} LQG. This dual loop model  corresponds to a quantum analog of teleparallel gravity since the dynamics is encoded in the torsion degrees of freeedom. As in LQG, we make a choice of representation for which the implementation of the kinematical constraint is particularly easy. We define in particular the \textit{group network} basis for the Hilbert space of translational invariant functionals. The group network basis which is labeled by SU(2) group variables diagonalizes the holonomy operators and the corresponding ground state is now peaked on flat geometries which are generically non-degenerate. Furthermore, the natural local excitations are now mass excitations which can be interpreted as spinless particles. In contrast to LQG, the ground state is not normalizable anymore.

Both quantization schemes therefore complement one another. The usual scheme is natural if one wants to extract a notion of quantum geometry. The dual picture seems more natural if one wishes to understand the coupling to massive particles. The nature of the duality relation that exchanges the corresponding bases will be detailed in a future work \cite{inprogress}. The fact that there are two complementary descriptions follows from a fundamental dichotomy inherent to the definition of geometry. On the one hand geometry can be characterized as the measurement of geometrical quantities, encoded in the tetrad which is an {\it electric} field. On the other hand, the geometry can be characterized via the notion of parallel transport which is encoded into the connection playing the role of {\it magnetic } field. From this perspective, the duality between the general relativity picture and the teleparallel picture can be understood as an electric-magnetic duality as studied for instance in the condensed matter literature \cite{Buerschaper2010}.\footnote{More specifically, the two bases associated with these quantization schemes are related to string net models built upon the category ${\rm Rep}[G]$ of finite dimensional representation and the category ${\rm Vec}_G$ of $G$-graded vector spaces, respectively.}

\bigskip \noindent
Of course the idea that there are two opposite representations of 3d gravity related by a duality that exchange curvature excitations with torsion excitations is not new. It has been investigated at length in a series of works  \cite{Freidel:2004vi,Freidel:2004nb} where it was shown that the Ponzano-regge model which implements the dynamics when working with the spin-network basis also carries curvature excitations and in fact representations of the Drinfel'd double of the gauge group. 
Using these results, excitations which can carry both mass and spin have been represented in the spin-network framework.
Furthermore, the explicit form of the duality transformation between the electric and the magnetic picture has also been expressed for simple graphs as a generalization of the Fourier transform in \cite{Freidel:2006qv}.
These results have been recently revisited in \cite{DDR1, DD16, Delcamp:2016eya, Delcamp:2017pcw} where an alternative gauge invariant basis was introduced for lattice gauge theories, namely the so-called \emph{fusion basis}, labeled by irreducible representations of the Drinfel'd double, which accounts for both curvature and torsion excitations.  Consequently, our work goes in line with  the ongoing efforts to propose alternative bases for LQG and more generally for topological quantum field theories (TQFTs) with defects \cite{ocneanu1993, ocneanu2001, KKR, Lan2013, DDR1, DD16, Dittrich:2017nmq, Delcamp:2017pcw}.  In this context, the ground state is typically peaked on flat connections, this is the so-called $BF$ vacuum \cite{DGfluxC, DGfluxQ}.

Note  that  since  we impose first the translational symmetry, a non-compact symmetry, we have two different possible quantizations available. The one considered in \cite{DGfluxC, DGfluxQ} is based on a Bohr compactification so as to obtain a finite scalar product between states, i.e. for example $\langle g_1|g_2\rangle = \delta_{g_1,g_2}$  where we use  the \textit{Kronecker} delta symbol. The other one, the  standard quantization, will involve distribution-valued scalar product, i.e. $\langle g_1|g_2\rangle = \delta_{g_1,g_2}$  where we use  the \textit{Dirac} delta function.
In the present paper, we want to quantize 3d gravity by imposing the flatness constraint first and using the standard quantization scheme (as opposed to Bohr's). We will naturally  obtain  a new basis labeled by group variables which is suited to study curvature excitations. To achieve this, we mimic the standard procedure that is used in the LQG case. Furthermore, we will provide what stands for the analogue of the Ponzano-Regge model (PR) \cite{Ponzano_Regge_1969}. 

Interestingly, the corresponding path integral quantization leads to a state-sum related to the Dijkgraaf-Witten (DW) \cite{dijkgraaf1990} model, which is usually defined in terms of finite groups.  It turns out that the Dijkgraaf-Witten model has been under intensive investigation in the context of topological phases with defects. For instance, large classes of (2+1)d and (3+1)d Levin-Wen models are defined as lattice Hamiltonian realizations of the 3d and 4d Dijkgraaf-Witten models \cite{hu2013twisted, Bullivant:2017qrv, 2017arXiv170404221L, Wang:2016rzy, Wan:2014woa, Delcamp:2018wlb}. As such, we will see how we can borrow some results from the topological order literature

\bigskip \noindent
The paper is organized as follows.  In sec.~\ref{sec:first}  we recall basic facts about 3d gravity, its discretization and a general overview of its quantization procedure. In sec.~\ref{sec:second}, we expose the quantization in the LQG scheme, which consists in implementing the Gau{\ss} constraint first, and emphasize the connection with the Ponzano-Regge model. We then present in sec.~\ref{F-G} the quantization when imposing the flatness constraint first, which we call the dual quantization scheme. We also recall in this section the construction of the Djkgraaf-Witten model for finite groups and explain how our dual quantization scheme is related to it. 

\section{3d gravity}
\label{sec:first}

\subsection{Continuum case}

In 2+1 dimensions, general relativity is a topological field theory with no local degrees of freedom known as $BF$ theory \cite{horowitz1989}. A first order action for general relativity with zero cosmological constant is therefore provided by
\be
	S[e,\omega] = \int_{\mathcal{M}}\text{tr}(e \wedge F(\omega))
\ee 
where $\mathcal{M} = \Sigma \times \mathbb{R}$, $e$ denotes a $\mathfrak{su}(2)$-valued 1-form, $\omega$ a connection on a trivial SU(2)-bundle and $F=\rd \omega +\omega \wedge \omega$ its curvature. Varying the action, we obtain the equations of motion:
\begin{align}
	\frac{\delta S}{\delta e} \, : \q  F(\omega)=0,  \q  \q
	\frac{\delta S}{\delta \omega} \, : \q  \text{d}_{\omega}e=0 
\end{align}
which are the defining equations for flat and torsionless connections. The first order action for gravity possesses two kinds of gauge symmetries. First, we have a local SU(2) rotation symmetry
\begin{equation}
	\delta_\Lambda e = [e,\Lambda] \q , \q \delta_\Lambda  \omega = {\rd}_{\omega} \Lambda
\end{equation}
with $\Lambda$ a $\mathfrak{su}(2)$-valued 0-form. Then, we have a translational symmetry parametrized by a $\mathfrak{su}(2)$-valued 0-form $N$
\be
	\delta_N  e = {\rd}_{\omega}N \q , \q \delta_N  \omega = 0
\ee
which is a consequence of the Bianchi identity ${\rd}_{\omega}F=0$. In order to apply Dirac's quantization program, we parametrize the phase space by the pull back of  the connection  $\omega$ and the coframe $e$  to $\Sigma$. We denote by $A^i_a$ the pull-back of $\omega$ and $E^b_j=\epsilon^{bc} \eta_{ji} e_a^i$ its conjugate variable in local coordinates. Their Poisson brackets read
\be
\{\, A^i_a(x)\, ,\, E_j^b(y)\, \} = \delta^i_j\delta_a^b\delta^{(2)}(x,y) \; .
\ee 
Canonical analysis of the action then reveals the following equations:
\be
	D_b E^b_j = 0 \q , \q  \q F^i_{ab}(A) = 0 \; ,
\ee
which are the first class constraints generating the local symmetries of the action. These two constraints are  referred to as the Gau{\ss} constraint $\mathcal{G}$ and the flatness constraint $\mathcal{F}$, respectively.

\subsection{Discretization}
Upon quantization, we need to choose a set of basis phase space functions which are then promoted to operators \cite{Ashtekar1991, Ashtekar1993, Ashtekar:1994mh}. However, in order to make such quantization feasible, we require two conditions on the basis: $(i)$ Poisson brackets between canonical variables which form an algebra and $(ii)$ that they possess simple expressions under  gauge transformations. So far, we have a phase space parametrized by $A^i_a$ and $E^b_j$ whose Poisson brackets are distributional. Furthermore, the $\mathfrak{su}(2)$-valued connection transforms as
\be
	g \triangleright A_a = gA_ag^{-1}+ g \partial_a g^{-1} \; .
\ee 
{
A simple way to achieve both $(i)$ and $(ii)$ is to consider holonomies of the connection $A$ along paths in $\Sigma$. More precisely, let $\gamma$ be a piecewise analytic curve, the holonomy $h_\gamma(A) \in {\rm SU(2)}$ along $\gamma$ in $\Sigma$ is given by the path ordered exponential
\be
	h_{\gamma}(A) = \mathcal{P}{\rm exp}\Big(\int_\gamma A \Big)
\ee
which transforms as
\be
	\label{Gauss_hol}
	g \triangleright h_\gamma =  g_{t(\gamma)} h_\gamma g_{s(\gamma)}^{-1} \; ,
\ee
where $s(\gamma)$ and $t(\gamma)$ denote the source and target nodes of $\gamma$, respectively. 
Similarly, the frame field is smeared over a one dimensional submanifold so as to define the flux variables

\be
	X_\gamma = \int_{e}h^{-1}_{\gamma,x}{\bf e}(x)h_{\gamma,x}{\rd}x
\ee
where $e$ intersects transversally $\gamma$ in one point, $h_{\gamma,x}$ is the holonomy going from the point $s(\gamma)$ to $x\in e$, and 
${\bf e}$ is the dyad. 
The flux variables transform as 
\be
	g \triangleright X_\gamma = g_{s(\gamma)}X_\ell g_{s(\gamma)}^{-1}
\ee
and the holonomy-flux algebra finally reads
\be
	\label{Poisson}
	\{h_\gamma,h_\gamma\} = 0 \q , \q \{X^a_\gamma,h_\gamma\}=h_\gamma \tau^a \q , \q \{X^a_\gamma , X^b_\gamma \} = \epsilon^{ab}{}_cX^c_\gamma \; ,	
\ee
with $\tau^a$ being the generators of $\mathfrak{su}(2)$ satisfying the algebra $[\tau^a,\tau^b]=\epsilon^{ab}{}_c\tau^c$. So far, we have been discussing the discretization of the phase spaces variables on a single curve $\gamma$. The next step consists in considering a full lattice such that each edge/link of this lattice carries a pair of discretized variables as previously described. 

In \cite{Dupuis:2017otn}, two different discretizations of the $BF$ action were proposed by identifying two different sets of variables which satsify the holonomy-flux algebra. On the one hand, the standard LQG formalism was recovered with holonomies living on the links $\ell$ of the 1-skeleton of the graph dual to a triangulation. The fluxes are then attached to the nodes of this 1-skeleton and depend on the triangulation edge $e$ dual to $\ell$, i.e.
\be
	\label{Poisson l}
	\{h_\ell,h_\ell\} = 0 \q , \q \{X^a_\ell,h_\ell\}=h_\ell \tau^a \q , \q \{X^a_\ell , X^b_\ell \} = \epsilon^{ab}{}_cX^c_\ell \; .
\ee
By construction, the fluxes satisfy Gau\ss \,  constraints which are associated with the nodes of  the 1-skeleton. This choice is natural when the dynamics is encoded in the flatness constraint.

On the other hand, it was also found a discretization where the holonomies now live on the edges $e$ of a discretization, whereas the fluxes are attached to its vertices and depend on the link $\ell$ dual to $e$, i.e. 
\be
	\label{Poisson e}
	\{h_e,h_e\} = 0 \q , \q \{X^a_e,h_e\}=h_e \tau^a \q , \q \{X^a_e , X^b_e \} = \epsilon^{ab}{}_cX^c_e \; . 	
\ee
In this case, the flatness constraint is naturally implemented around the nodes of the 1-skeleton. This choice is more natural when the dynamics is encoded in the Gau\ss \, constraint. We study in this manuscript the two different canonical quantizations of (2+1)d general relativity naturally associated with these two discretizations of $BF$ theory.
}

\subsection{Quantization \`a la Dirac \label{sec:quantDirac}}

Roughly speaking, Dirac's quantization program consists in: $(i)$ Choosing a representation of the phase space variables as operators in a so-called \emph{kinematical} Hilbert space $\mathcal{H}^{\rm kin}$, $(ii)$ promote the constraints to operators in this Hilbert space, $(iii)$  finally find the states solutions of these quantum constraints. The space of solutions together with the physical inner product define the physical Hilbert space $\mathcal{H}^{\rm phys}$. 

What makes gravity special when formulated in terms of the frame fields ${e}_a= {e}_a^\mu \partial_\mu$ (with $a$ internal indices) is the fact that its symmetry group is the product of local gauge transformations and diffeomorphisms.
We can then choose one of the symmetries to be the kinematical one and  implement the other one dynamically. 
The conventional choice is to choose the local gauge transformations to be kinematical. This means that we work in metric variables $g_{\mu\nu}= \eta_{ab}e_{\mu}^a e_\nu^b$ for which gauge symmetry is trivially implemented while diffeomorphism invariance is the  defining property of the action $S(g)$.
The other choice is to work in terms of torsion variables $ T_{ab}{}^c= e^c([{e}_a,{e}_b])$, which is a scalar under diffeomorphism, and then implement the gauge symmetry non trivially into the choice of an action $S(T)$ which is gauge invariant.  In this second option, diffeomorphism invariance is implemented kinematically while gauge invariance is implemented dynamically. In this paper, we exploit the fact that the corresponding constraints, namely the Gau{\ss} constraint and the flatness constraint, are implemented one after the other in a specific order. The choice of ordering goes together with a choice of representation. Depending on such a choice, one constraint will have a more natural action on the kinematical Hilbert space $\mathcal{H}^{\rm kin}$.

\begin{figure}[t]
	\begin{center}
	\includegraphics[scale=1]{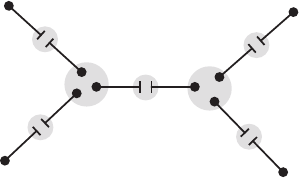}
	\caption{Basis states for the kinematical Hilbert space are obtained by first gluing half-link (resp. half-edge) states so as to obtain the link (resp. edge) states, and then gluing together half-links (resp. half-edges) meeting at a given node (resp. vertex). Both types of gluing involve the kinematical constraint which requires us to choose a representation for the half-link (resp. half-edge) states on which this contraint acts naturally.
	}\label{glueing}
	\end{center}
\end{figure}

\medskip
\noindent
The definition of the kinematical Hilbert space $\mathcal{H}^{\rm kin}$ follows several steps. First we need to pick a graph which can be either the one-skeleton of a discretization or its dual graph. In the former case we will refer to the 0-simplices, 1-simplices and 2-simplices as vertex, edges and faces while in the latter one we will talk about nodes, links and plaquettes. Each node (resp. vertex) has incoming and outgoing half-links (resp. half-edges) which are associated with a given state. These half-links (resp. half-edges) need then to be \emph{glued} together so as to form full link (resp. edge). Furthermore, half-links (resp. half-edges) meeting at a node (resp. vertex) also need to glued together, see fig. \ref{glueing}. The gluing conditions are obtained by putting some constraints on the corresponding states, which in turn give rise to the notion of \textit{fusion tensor product} denoted by $\boxtimes$, i.e. a modified tensor product taking into account the constraints.

Such fusion tensor product can be seen as a generalization of the usual notion of tensor product \cite{Thom,Wassermann,Gaberdiel:1993td}. The usual tensor product between two vector spaces $V$ and $W$ is actually an equivalence class under the action of a ring $\mathbb K$ (most often $\C$) such that
\bes
	\label{tensor}
	v\otimes w\in V\ot_{\mathbb K} W \q  \textrm{ such that } \q ( va )\,\ot \, w\, &\sim&\, v \ot \, (a w ) \, , \, a\in {\mathbb K} \; .
\ees
Studies from conformal field theories called for a generalization of this structure \cite{Wassermann,Gaberdiel:1993td}. In this context, one considers two Hilbert spaces, carrying some representation of some non-abelian group (such as the conformal group), associated with spacetime points. In the limit where two of these spacetime points coincide, the corresponding  Hilbert spaces have to "fuse" such that the (non-abelian) transformations associated with each point should match. This is precisely a non-commutative generalization of the defining property of a tensor product \eqref{tensor}, where $a$ would now belong to a non-commutative ring. The same strategy applies in the context of gluing half-links (resp. half-edges) in order to construct the kinematical Hilbert space. 

In the following, we will be interested in two different symmetry groups $G$, namely the group of translations and the group of rotations. The kinematical Hilbert space associated with a full link (resp. edge) $\cH^{\rm kin}$ will then be defined as  the fusion product of Hilbert spaces $ \cH_{L,R}$ associated with the corresponding left and right half-links (resp. half-edges), i.e. $ \cH^{\rm kin} \sim \cH_L \boxtimes_G \cH_R$. Remark that the Hilbert spaces under consideration should be bimodules themselves since we require to have a group action at each one of the extremities of the half-links (resp. half-edges).  Starting from the bimodule $\cH_L \otimes \cH_R$ which does posses a left and a right action of the symmetry group $G$ associated with each half-link (resp. half-edge), we define the corresponding fusion tensor product as
\be
	\label{fusion}
	\cH_L \boxtimes_G \cH_R \ni v \boxtimes w \q \textrm{ such that }   \q v \boxtimes \, ( w\triangleleft g)   \sim\, (g \triangleright v )\,\boxtimes \, w \, , \, g\in G \; .
\ee
In this sense, we are gluing the half-links (resp. half-edges) via a 2-leg intertwiner\footnote{Note that provided the symmetry group $G$ is equipped with a left (or right) Haar measure, the equivalence class can be represented as 
\be\label{fusion bis}
	v \boxtimes w \equiv \int \rd  t \, (t \triangleright v) \ot (w \triangleleft t). 
\ee}
(between bimodules). Since equivalence classes are defined with respect to a given symmetry, which in turn is associated with a given constraint, this gluing step encodes the implementation of some constraint. In our context, this constraint is the so-called kinematical one which can be either the Gau{\ss} constraint or the flatness constraint according to the quantization scheme under consideration. It turns out that such construction naturally gives rise to (possibly maximal) entanglement between fundamental states \cite{Donnelly:2016auv}. This is a possibe way to interpret entanglement as the ``fabric of space(-time)'' \cite{eugenio}.

\bigskip \noindent
More specifically, in LQG, we usually choose the holonomies as the configuration variables. We then pick a graph $\Gamma$ embedded on $\Sigma$ dual to a discretization and define the kinematical Hilbert space $\mathcal{H}_\Gamma^{\rm kin}$ as the space of square integrable functionals of holonomies defined along the links of $\Gamma$. Furthermore, we choose to implement the Gau{\ss} constraint, which acts at the endpoints of the holonomies according to \eqref{Gauss_hol}, at the kinematical level. In order to make the implementation of the Gau{\ss} constraint easy, we choose a basis in which its action is natural. This is the so-called \emph{spin-network} basis for which the constraint is imposed via intertwiners between irreducible representations of SU(2). 
In the spin network basis, the {\it links} $\ell$ of the graph $\Gamma$ are the support of the holonomies $\{g_\ell\}$, while the {\it nodes} $n$ of $\Gamma$ are the support of the action for the flux operators $\{X_n\}$. Imposing gauge invariance at the nodes forces the vector associated with each node to be an intertwiner. It is customary to choose the graph $\Gamma$ to be trivalent in order to make such intertwiners uniquely defined. The choice of basis acts as a choice of parametrization for the states. Such a choice will persist when implementing the flatness constraint which generate the dynamics of the theory.

We propose in this paper an alternative quantization procedure for $BF$-theory which relies upon switching the order of implementation of the constraints. The flatness constraint is therefore implemented at the kinematical level. This goes together with an alternative choice of representation. Indeed, the flatness constraint has a more natural action directly in terms of holonomies. The Gau${\ss}$ constraint will then encode the dynamics of the theory. We refer to this alternative procedure as  dual LQG, denoted by LQG$^\star$. 

In this alternative picture we work in terms of a graph $\Gamma^\star$.
The {\it edges} $e$ of $\Gamma^\star$ supports the flux operators $X_e$ while the action of the translation operator is supported at the {\it vertices} $v$ of $\Gamma^\star$.
In order to mimic the simplicity of the spin network representation, we will work 
out the detail of the construction for dual graph $\Gamma^\star$ which are trivalent.
Such a graph can be thought as the one-skeleton $\hex_1$ of a polytope decomposition $\hex$ with only trivalent vertices. The flatness constraint is therefore implemented at the $0$-cells. In terms of the original graph, this would correspond to a situation where the flatness constraint is implemented at triangular plaquettes $p \subset \Gamma$  involving only three holonomies.  These different manipulations are depicted in tab.~\ref{summary}.

\begin{table}[t]
	\center
	\begin{tabular}{ l|c|c } 
		\q & LQG & LQG$^\star$ \\
		\hline\hline 
		\begin{tabular}{l} 
			Graph $\Gamma$ \\[-0.5em]
			\q- nodes $n$ \\[-0.5em]
			\q- links $\ell$ \\[-0.5em]
			\q- plaquettes $p$
		\end{tabular} 
		& $	\includegraphics[scale=1, valign=c]{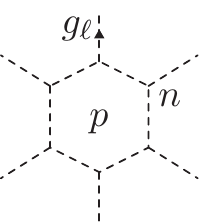}$ 
		& $\includegraphics[scale=1, valign=c]{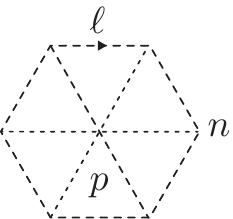}$ \\[3em]
		\begin{tabular}{l} 
			Discretization \\[-0.5em]
			\q- vertices $v$ \\[-0.5em]
			\q- edges $e$ \\[-0.5em]
			\q- faces $f$
		\end{tabular} 
		& \begin{tabular}{c}
			$\includegraphics[scale=1, valign=c]{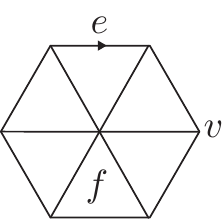}$ \\[0.5em]
			\q\q\; Triangulation $\large\triangle$ \q\q\;
		\end{tabular}
		& \begin{tabular}{c}
			$\phantom{(}$ \\[-1.8em]
			$\includegraphics[scale=1, valign=c]{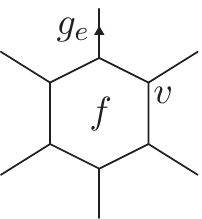}$ \\[-0.3em]
			Polytope decomposition $\large\hexagon$
		\end{tabular} \\[4em]
		Kinematics
		& $\mathcal{G}_n = \displaystyle\sum_{\ell \supset n}^{\rightarrow}g \triangleright X_{\ell} \overset{!}{=}0$
		&$\mathcal{F}_v^\star = \displaystyle\prod_{e \supset v}^{\rightarrow}g_e \overset{!}{=} \mathbbm{1}$  	 \\[3em]
		Dynamics
		& $\mathcal{F}_p = \displaystyle\prod_{\ell \subset p}^{\rightarrow}g_{\ell} \overset{!}{=} \mathbbm{1}$
		& 	$\mathcal{G}_f^\star = \displaystyle\sum_{e \subset f}^{\rightarrow}g \triangleright X_{e} \overset{!}{=}0$\\
	\end{tabular}
	\caption{Summary of the construction.}\label{summary}
\end{table}
\noindent
To summarize, we represent the holonomies on the edges $e$ of the one-skeleton $\hex_1$ of a polytope decomposition $\hex$ such that the flatness constraint is enforced at the vertices $v$ of this one-skeleton. Thus, we effectively end-up with a configuration very similar to the usual one at the difference that the roles of the Gau{\ss} constraint and the flatness constraints are exchanged. The two choices of ordering described above yields the two quantization schemes for 3d general gravity we consider in this paper. These two schemes can be summed up as follows:
\begin{align} \nn
	&\,{\rm LQG} \; : \; \mathcal{H}^{\rm kin}_\Gamma \; \longrightarrow 
	\; \mathcal{H}_\Gamma^\mathcal{G} \; \longrightarrow \;
	\mathcal{H}_\Gamma^{\rm phys}\\ \nn
	&{\rm LQG^\star}  : \; \mathcal{H}^{\rm kin}_{\hexagon_1} \; \longrightarrow 
	\; \mathcal{H}_{\hexagon_1}^\mathcal{F} \!\! \longrightarrow \;
	\mathcal{H}_{\hexagon_1}^{\rm phys}
\end{align}
where $\mathcal{H}_\Gamma^\mathcal{G}$ and $\mathcal{H}_{\hexagon_1}^\mathcal{F}$ refers to the space of kinematical states satisfying the Gau{\ss} constraint or the flatness constraint, respectively. It is worth emphasizing one more time that this choice of ordering of the constraints, and {\it a fortiori} choice of basis for the kinematical Hilbert space, has important consequences. Indeed, choosing a parametrization determines a preferential type of excitations for the theory, these excitations being defined with respect to a given vacuum. We will see that, if the usual spin network basis is particularly adapted to torsion excitations, the basis used in ${\rm LQG^\star}$ naturally encodes curvature excitations.

\section{Canonical quantization: $\mathcal{G} \rightarrow \mathcal{F}$}
\label{sec:second}

In this section, we briefly review well-known aspects of the derivation of Loop Quantum Gravity in (2+1)d \cite{rovelli2004, Perez:2004hj, thiemann}. In particular, we emphasize how the implementation of the flatness constraint is performed by the Ponzano-Regge state-sum \cite{Ponzano_Regge_1969, Barrett:2008wh} using the language of Levin-Wen models \cite{Levin:2004mi}. This will be used as a comparison point against the new quantization we propose in the following subsection.

\subsection{Kinematical space and representation of the holonomy-flux algebra}

In the standard picture of LQG, the Hilbert space $\cH_\Gamma^{\mathcal{G}}$ is built from spin network states, which naturally solve the Gauss constraint. 
We will now recall the main steps which lead to the definition of such basis states. First, the Poisson brackets \eqref{Poisson l} are turned into commutators so as to obtain the \emph{holonomy-flux} algebra  $\cA_\Gamma$ associated with the graph $\Gamma$. This algebra is a direct sum of \emph{link algebras} $\cA_\ell$ associated with each link $\ell \subset \Gamma$ and generated by the pair $(\widehat{X}^j_\ell, \widehat{h}_\ell)$ satisfying the commutation relations
\be
	\label{loopalg}
	[\widehat{h}_\ell,\widehat{h}_\ell] = 0 \q , 
	\q [\widehat{X}^a_\ell ,\widehat{h}_\ell] = i \widehat{h}_\ell \tau^a \q , \q  
	[\widehat{X}^a_\ell , \widehat{X}^b_\ell ] = i  \epsilon^{ab}{}_{c}\widehat{X}^c_\ell \; .
\ee
At this stage, several important facts should be noticed in order to prepare for the dual quantization scheme. Firstly, we note that this algebra contains two sub-algebras: A non-commutative algebra which is generated by $\widehat{X}^a_\ell$ and a commutative one which is generated by the matrix element operators $\widehat{h}_\ell$. Secondly, we remark that the combination $\widehat{X}_{\ell^{-1} }:= -h_\ell \widehat{X}_{\ell}h_\ell^{-1}$ commutes with $\widehat{X}_{\ell}$ while satisfying the same commutation relations   $[\widehat{X}_{\ell^{-1}}^{ a},\widehat{X}_{\ell^{-1}}^{b}]= i \epsilon^{ab}{}_c \widehat{X}_{\ell^{-1}}^c$.
The fact that $\widehat{X}_{\ell}$ and $\widehat{X}_{\ell^{-1}}$ commute follows from the property that $\widehat{X}_{\ell}$ acts as the left invariant derivative on functions of the holonomy
while $\widehat{X}_{\ell^{-1}}$ acts as a  right invariant derivative: $[\widehat{X}_{\ell^{-1}}^{a}, \widehat{h}_\ell]= -i \tau^a \widehat{h}_\ell$. 
Under reversal of the orientation we also assume that $\widehat{h}_{\ell^{-1}}= \widehat{h}_\ell^{-1}$. This implies in particular that the algebra $\cA_\ell$ is independent  on the choice of orientation of the edge.

The choice of a representation of the algebra $\cA_\ell$ is characterized by a choice of maximally commuting sub-algebra. Any maximally commuting algebra is three-dimensional and there are two natural choices for this sub-algebra. The first choice amounts to diagonalizing the set of fluxes  
\be\label{maxcom}
	\widehat{X}_\ell^2=  \widehat{X}_{\ell^{-1}}^2 \q , \q \widehat{X}_\ell^3 \q , \q \widehat{X}_{\ell^{-1}}^3 \; .
\ee 
This is the choice made in the construction of the usual LQG basis since it is well-adapted to the case where we solve first the Gau{\ss} constraint which is specified in terms of the fluxes $\widehat{X}_{\ell} $ meeting at a node $n$. We will therefore focus on the non-commutative sub-algebra generated by the fluxes for now.
Note already that in the dual picture, it will be more natural to focus on the other choice, namely the holonomy sub-algebra which is generated by holonomy operators $\widehat{h}_\ell$. This will correspond to the case where we impose kinematically the flatness constraint.

The representation that diagonalizes the flux operators (\ref{maxcom}) is labeled by  $\su(2)$-irreducible representations $V_j$. Due to the nature of the commuting operators, we expect that the link Hilbert space $\cH_\ell$ to be characterized by the representations $j_{\ell}$, $j_{\ell^{-1}}$ together with the corresponding magnetic numbers. By construction, we have the constraint
$|\widehat{X}_\ell|= |\widehat{X}_{\ell^{-1}}|$ which in turn imposes  $j_{\ell}=j_{\ell^{-1}} \equiv j$. This suggests that the natural Hilbert space for a link $\ell$ is
\be
	\label{Hlink}
	\cH_\ell\equiv \bigoplus_{j} V_{j}\otimes V_{j^\ast} \ni |j ,m,n \ra \equiv | j, n \ra \la j ,m |\
\ee
where we use the dual representation $V_{j^\ast}$ for the right-hand side in order to keep track of the orientation of the link. It is understood from the notation that the links are oriented such that the magnetic numbers $m$ and $n$ are associated with the target and source nodes, respectively. We will refer to the condition $|\widehat{X}_\ell|= |\widehat{X}_{\ell^{-1}}|$ as the \emph{matching condition}. The implementation of such condition can be made more explicit by defining the projector $\mathfrak{P}: |j,n \ra \la j',m| \mapsto \delta_{jj'}|j,n\ra \la j,m|$. We will make use of such a projector later on. Furthermore, we remark that the Hilbert space $\cH_\ell$ is actually an $\su(2)$-bimodule since the fluxes $\widehat{X}_\ell$ acts as left invariant derivatives while the fluxes $\widehat{X}_{\ell^{-1}}$ act as right invariant derivatives. These correspond to infinitesimal generators of right and left translations, respectively, i.e.
\be
	\label{action flux}
	\widehat{X}^a_\ell |j,m,n \ra = i  \sum_p |j,m,p \ra D^{j}_{pn}(\tau^a)
	\q , \q
	\widehat{X}^a_{\ell^{-1}} |j,m,n \ra = i  \sum_q |j,q,n \ra D^{j}_{mq}(-\tau^a)
\ee
so that the spaces spanned by $\{\,|j,m,p \ra \; | \; p=-j,\dots,+j\}$ and $\{\,|j,q,n \ra \; | \; q=-j,\dots,+j\}$ are sub-representation spaces respectively carrying a representation $D^{j}$ and a contragradient representation $D^{j^\ast}$. 

In order to have a complete picture, we also need to identify the action of the holonomy operator, or more exactly of the matrix element operators $\widehat{h}_{BA}$. For simplicity, we choose to express it in the spinor representation. Defining the Clebsch-Gordan coefficients $C^{\, j_1 \; j_2 \; j_3}_{m_1 m_2 m_3}$ via
\begin{equation} 
	\sum_{m_1,m_2} C^{\, j_1 \; j_2 \; j_3}_{m_1m_2m_3} 
	|j_1,m_1\ra \otimes |j_2, m_2\ra =
	|j_3,m_3\ra \; ,
\end{equation}
and similarly its conjugate  $\overline{C}^{\, j_1 \; j_2 \; j_3}_{m_1m_2m_3}$  via
\begin{equation}
	|j_1,m_1\ra \otimes |j_2, m_2\ra = \sum_{j_3,m_3}
	 \overline{C}^{\, j_1 \; j_2 \; j_3}_{m_1m_2m_3} |j_3,m_3\ra \; ,
\end{equation}
the action of the holonomy matrix element operators reads
\be 
	\label{HolonomyAction}
	\widehat{h}_{BA} \ | j, m, n\ra 
	 = \sum_{J = j\pm \frac{1}{2}} \sum_{-J \leq M\leq J \atop -J \leq N\leq J}  \overline{C}^{\, \frac{1}{2}  \, j \, J}_{A \, n N}\, {C}^{\frac{1}{2} \; j \; J}_{B m M} | J, M,N\ra
\ee
where $\widehat{h}_{BA}=\la B |\widehat{h}|A\ra$ and $|A\ra=|1/2,A\ra$ simply denotes the state $A$ in the spinorial representation.

\bigskip \noindent
The definition \eqref{HolonomyAction} shows that the holonomy operator acts on both side of the link, unlike the flux operator.\footnote{This could also be illustrated   using the graphical calculus introduced below (see \eqref{jj}).} As a consistency check, we can show how it is possible to recover a representation of the link algebra \eqref{loopalg}. To do so, we first rewrite the action \eqref{HolonomyAction} using the definition \eqref{Hlink} together with the projector $\mathfrak P$:
\be
	\widehat{h}_{BA} \ | j, m, n\ra= \sum_{J = j\pm \frac{1}{2}} \sum_{-J \leq M\leq J \atop -J \leq N\leq J} 
	\overline{C}^{\, \frac{1}{2}  \, j \, J}_{A \, n N} \, |J,N\ra \la J,M| \, 
	{C}^{\frac{1}{2} \; j \; J}_{B m M}  
	={\mathfrak P} \big( \, |A\ra\otimes |j,n \ra \la j,m| \otimes \la B| \, \big) 
\ee
where $\la j,m| \otimes \la B| := (|B\ra \otimes | j,m\ra)^\dagger$.
We can now evaluate the following quantity\footnote{ In order to establish this, one uses the fact that 
	$\widehat{X}^a |j,n\ra \la j,m|  = D^j(\tau^a)|j,n\ra \la j,m| $ 
	 together with the group action \eqref{haction} so that 
	 \bes
		\widehat{X}^a \widehat{h}_{BA} |j,n\ra \la j,m| &=& {\mathfrak P} \big([\tau^a\otimes 1 +1\otimes D^j(\tau^a)]  \, |A\ra\otimes |j,n\ra \la j,m| \otimes \la B| \, \big)\cr
		\widehat{h}_{BA} \widehat{X}^a  |j,n\ra \la j,m| &=& {\mathfrak P} \big([1\otimes D^j(\tau^a)]  \, |A\ra\otimes |j,n\ra \la j,m| \otimes \la B| \, \big) \; .
	\ees
}
\bes
	[\widehat{X}^a,\widehat{h}_{BA}] \ | j, m, n\ra = 
	{\mathfrak P} \big(\tau^a \, |A\ra\otimes |j,n\ra \la j,m| \otimes \la B| \, \big)  = 
	\Big(\sum_{A'}\tau^a_{A'A}  \widehat{h}_{BA'}\Big) | j, m, n\ra 
\ees
and we recover the expected commutation relation between $\hat{h}$ and $\widehat{X}$.
We can also express the action of $h^{-1}_{AB}$ as 
\be \label{haction}
	(\widehat{h}^{-1})_{AB} \ | j, m, n\ra
	=	{\mathfrak P} \big(\tau^a \, |A^\ast \ra\otimes |j,n\ra \la j,m| \otimes \la B^\ast| \, \big)
\ee
where $|A^*\ra= (-1)^{1/2-A}|-A\ra$ is the conjugate state.
The identity $\sum_A\widehat{h}_{BA} (\widehat{h}^{-1})_{AC} =\delta_{BC}$ finally follows from the fact that 
$\sum_A |A\ra\otimes |A^\ast\ra$ is the singlet state.

\bigskip \noindent
The next step is to precise what is the Hilbert space structure associate with a link $\ell$. 
From the fact that the states $|j,m,n\ra$ diagonalizes $X^3_\ell,X_{\ell^{-1}}^3$ and $(X_\ell)^2$, we already know that they form an orthogonal basis. Demanding that 
$(\widehat{h}_{BA})^\dagger = (\widehat{h}^{-1})_{AB}$ then forces the normalization condition 
\be
	\la j',m',n'|j,m,n\ra = \frac{1}{d_j} 
	\delta_{jj'} \delta_{mm'} \delta_{nn'} 
\ee
where $d_j= 2j+1$.
Moreover, thinking of the states $|\phi\ra=\sum_j \phi_j$, with 
$\phi_j=\sum_{mn} \phi_{jmn} |j,m,n\ra$, as endomorphisms $\widehat{\phi} =
\sum_{jmn} \phi_{jmn} |j,n\ra\la j,m|$, we can express the previous scalar product as a weighted trace: 
\be
	\label{Trace}
	\la \phi |\psi\ra = {\rm tr} (\widehat{\phi}^\dagger \widehat{\psi}) :=\sum_j \frac1{d_j} \tr_{V_j}(\widehat{\phi}_j^\dagger \widehat{\psi}_j) \; .
\ee
Furthermore, it follows from the symmetry property\footnote{ Explicitely given by 
\be
	\frac1{\sqrt{ {\rd}_{j_3}}} C^{\, j_1 \; j_2 \; j_3}_{m_1m_2m_3} = \frac{(-1)^{j_1-m_1}}{\sqrt{ {\rd}_{j_2}}}
	C^{\, j_3 \; j_1 \; j_2}_{m_3m_1^* m_2} \q , \q 
	 C^{\, j_1 \; j_2 \; j_3}_{m_1m_2m_3} =(-1)^{j_1+j_2-j_3} C^{\, j_2 \; j_1 \; j_3}_{m_2m_1m_3} \; .
\ee}
of the Clebsh-Gordan coefficients that
\begin{align}
	\nn
	\label{symCG}
	\la J,M,N|\widehat{h}_{BA}|j,m,n\ra &=  \frac{1}{d_J} 
	\overline{C}^{\, \frac{1}{2}  \, j \, J}_{A \, n N}\, {C}^{\frac{1}{2} \; j \; J}_{B m M}
	=
	\frac{(-1)^{1-A-B}}{d_j} \overline{C}^{\;\; \frac{1}{2} \;  J \; j}_{-A N n}\, {C}^{\;\; \frac{1}{2} \; J \;\, j}_{-B M m}
	\\
	&=\la j,m,n|(\widehat{h}^{-1})_{AB}| J,M,N\ra \; .
\end{align}
\noindent
Now that we have the holonomy action on the spin states we can construct the  holonomy state $|g)$ which diagonalizes $\hat{h}$. This is the state that enters in the wave functional  $\psi(g)\equiv (g|\psi\ra $ and its explicit expression in terms of the states $|j,m,n \ra$ is provided by the generalized Fourier transform
\be
	\label{genFT}
	|g)= \sum_{j,m,n}{d_j} \overline{D^{j}_{ m n}(g)}|jm n\ra =  
	\sum_{jm n} {d_j} | j,n\ra \overline{D^{j}_{ m n}(g)} \la j,m | \; .
\ee
We can check that the action of the holonomy on such state is diagonal, i.e. $\widehat{h}_{BA} |g) =g_{BA}|g) $ as follows: 
\begin{align}
	\nn
	\sum_A\widehat{h}_{BA} |g)
	&=  \sum_{j,m,n}\sum_{J,M,N}
	d_j \overline{D^{j}_{mn}(g)}\,
	\overline{C}^{\, \frac{1}{2}  \, j \, J}_{A \, n N} \, |J,N\ra \la J,M| \, 
	{C}^{\frac{1}{2} \; j \; J}_{B m M}  \\ \nn
	&=  \sum_{j,m,n }\sum_{J,M,N} \sum_{B', n', N'}
	d_j D^{j}_{nm}(g^{-1})\,
	\overline{C}^{\frac{1}{2} \,  j \, J}_{A \, n N} \, |J,N \ra \la J,M| \,
	D^{\frac{1}{2}}_{BB'}(g)D^j_{mm'}(g)\overline{D^J_{MM'}(g)} \, 
	{C}^{\, \frac{1}{2} \;\, j \;\, J}_{B' m' M'} \\ \nn
	&=  \sum_{j,n }\sum_{J,M,N} \sum_{A', M'}
	d_j 
	D^{\frac{1}{2}}_{BB'}(g)\overline{D^J_{MM'}(g)}\,
	\overline{C}^{\frac{1}{2} \,  j \, J}_{A \, n N} \, |J,N \ra \la J,M| \, 	{C}^{\, \frac{1}{2} \;\, j \; J}_{B' n \, M'}  \\
	&= \sum_{J,M,N} 
	d_J
	D^{\frac{1}{2}}_{BA}(g)\overline{D^J_{MN}(g)} \,
	|J,N\ra \la J,M|  = g_{BA} |g ) \; .
\end{align}
In the first line we used the formula \eqref{HolonomyAction} for the action of the holonomy operator. In the second line we made use of the property $\overline{D^j_{mn}(g)} = D^j_{nm}(g^{-1})$ together with the invariance property
\begin{equation*}
	\sum_{n_1,n_2,n_3}D^{j_1}_{m_1n_1}(g)D^{j_2}_{m_2n_2}(g)
	\overline{D^{j_3}_{m_3n_3}(g)}\,	C^{ j_1 \, j_2 \, j_3}_{n_1n_2n_3} = 
		C^{\, j_1 \; j_2 \; j_3}_{m_1m_2m_3} \; .
\end{equation*}
Finally we used a symmetry property similar to \eqref{symCG} in order to make us of the orthogonality of the Clebsch-Gordan coefficients
\begin{equation*}
	\sum_{j_3,m_3}{C}^{\, j_1 \;  j_2 \; j_3}_{m_1 m_2 m_3}
	\overline{C}^{\, j_1 \;  j_2 \; j_3}_{n_1 n_2 m_3} = \delta_{m_1n_1}\delta_{m_2n_2} \; .
\end{equation*}
It turns out that we could have defined directly the  Hilbert space $\cH_\ell$ as obtained from the space of functions on SU(2) after decomposition via the Peter-Weyl theorem. Nevertheless, it was important to present carefully the steps behind its construction in order to introduce the dual construction later on. Furthermore, it is now possible to explain how such a representation arises from the general gluing procedure of half-links sketched in sec.~\ref{sec:quantDirac}.

\subsection{ Fusion tensor product and entanglement entropy}
As alluded earlier, we can apply the general procedure described in sec.~\ref{sec:quantDirac} in order to define the Hilbert space associated with a link from the fusion tensor product of the Hilbert spaces associated with the corresponding half-links.  

We start with two half-links referred to as left $\ell_L$ and right $\ell_R$ so that the corresponding half-link algebras are denoted by $\cA_{\ell_{L,R}}$ and the corresponding half-link Hilbert spaces by $\cH_{\ell_L,\ell_R}$, respectively. States living in such Hilbert spaces are denoted by 
\begin{align*}
	|j_L, m_L,n_L\ra &
	\equiv \includegraphics[scale=1, valign=c]{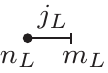} \in \cH_{\ell_L} \\
	|j_R,m_R,n_R\ra &
	\equiv \includegraphics[scale=1, valign=c]{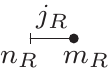} \in \cH_{\ell_R} \; . 
\end{align*}
Note that since half-links have two endpoints, these Hilbert spaces are actually bimodules. Our goal is to implement the fusion product so as to recover the Hilbert space $\cH_\ell$ for the full link, i.e.
\be
	\cH_\ell \simeq \cH_{\ell_L} \boxtimes_{\rm SU(2)} \cH_{\ell_R} \; .
\ee
To do so we need to look at the equivalence classes of states such that the action of the flux operator on the right of the left half-link and the one on the left of the right half-link are the same. We refer to this requirement as the \emph{matching constraint} which is directly related to the implementation of the Gau{\ss} constraint. We know that the action of the flux operator at one end of the half-link and its action at the other end are related via parallel transport. We denote by $\widehat{X}_L$ the flux operator acting on the left of $\ell_L$ and $\widehat{X}_R$ the flux operator acting on the right of $\ell_R$. The operators acting on the right of $\ell_L$ and on the left of $\ell_R$ are obtained via parallel transport as $\widehat{g}_L \triangleright \widehat{X}_L$ and $\widehat{g}_R \triangleright \widehat{X}_R$, respectively, so that the matching constraint explicitly reads
\be
	\widehat g_L \triangleright \widehat X_L 
	= \widehat g_R \triangleright \widehat X_R
\ee
which can be rewritten
\be
	(\widehat g_R ^{-1}\widehat g_L) \triangleright \widehat X_L 
	= \widehat X_R \; .
\ee
By identifying $\widehat{X}_\ell$ with $\widehat{X}_L$ and $\widehat{g}_\ell$ with $(\widehat g_R ^{-1}\widehat g_L)$, we finally obtain
\be
	\widehat h_\ell \triangleright \widehat X_L = \widehat X_R 
	\; \Longleftrightarrow \; 
	-\widehat X_{\ell^{-1}}= \widehat h_\ell\triangleright  \widehat X_{\ell} \; . 
\ee
This finally implies that the basis states for the Hilbert space $\cH_\ell$ are obtained from the gluing of half-links states as the following fusion tensor product
\be
	|j_L,m_L,n_L\ra  \boxtimes_{\rm SU(2)} |j_R,m_R,n_R\ra 
\ee 
which satisfies
\be
	\label{equivAction}
	 \big(|j_L,m_L,n_L\ra \triangleleft g \big) \boxtimes_{\rm SU(2)} |j_R,m_R,n_R\ra =  |j_L,m_L,n_L\ra  \boxtimes_{\rm SU(2)} \big( g \triangleright |j_R,m_R,n_R\ra \big) \; ,
\ee
where the right and left group actions read
\begin{align}
	g \triangleright |j_R,m_R,n_R\ra &\equiv\sum_q D^{j_R}_{m_Rq}(g^{-1}) \, |j_R,n_R\ra\la j_R , q |\\
	|j_L,m_L,n_L\ra \triangleleft g &\equiv \sum_p |j_L,p\ra\la j_L , m_L| \, D^{j_L}_{pn_L}(g) \; .
\end{align}
Equivalence \eqref{equivAction} implies that the fusion tensor product projects down to states that have matching $j_L$ and $j_R$, as well as matching $m_R$ and $n_L$, and on zero otherwise. This is nothing else than the definition of a bivalent intertwiner which implements the Gau{\ss} constraint at the bivalent node along which the gluing of the half-links is performed, i.e.
\begin{align}
	|j_L,m_L,n_L\ra  \boxtimes_{\rm SU(2)} |j_R,m_R,n_R\ra  &\sim   |j_L,n_L\ra \la j_L, m_L       |j_R,n_R\ra\la j_R ,m_R| \\
	&=  |j_L,n_L\ra \delta_{j_L,j_R} \delta_{n_R,m_L}  \la j_R, m_R| \; . 
\end{align}
We therefore recover the full link state depicted as
\be
	|j,n_L \ra \la j , m_R | \equiv 
		\includegraphics[scale=1, valign =c]{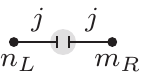}
\ee
where the gray dot represents the matching constraint.

\bigskip
\noindent
Remark that we could also have used insights from twisted geometries \cite{Freidel:2010aq} in order to construct this fusion product. Indeed, in such context the end-points of half-links that have to be fused are associated with a U(1) symmetry which corresponds to the Cartan subgroup of SU(2). Hence instead of considering $\textrm{SU}(2) \times \textrm{SU}(2)$ bimodules for the half-links, we could have considered the bimodule $\cH'_{\ell_L}$ with respect to the action 
$\textrm{SU}(2) \times \textrm{U}(1)$ for the left half-link and the bimodule $\cH'_{\ell_R}$ with respect to the action $\textrm{U}(1) \times \textrm{SU}(2)$ for the right half-link. Elements in these bimodules are then characterized by a representation of SU(2) and a singlet vector\footnote{Since U(1) is abelian, the irreducible representations are one dimensional.} which is a representation of U(1). The corresponding states are therefore denoted by 
\begin{align*}
	 |j_L, n_L\ra \la j_L, j_L| \in \cH'_{\ell_L} \q , \q  |j_R,j_R\ra \la j_R, m_R| \in \cH'_{\ell_R} \; .
\end{align*}
We then recover precisely the same bimodule states as before according to
\be
	\cH_\ell \simeq \cH'_{\ell_L} \boxtimes_{ \rm U(1)} \cH'_{\ell_R}
\ee
where the fusion tensor product simply identifies the spins $j_L$ and $j_R$. 
This construction allows to have a direct match with the twisted geometry framework. It could also provide new ways to consider the inclusion of massive particles since these are associated to the Cartan subgroup U(1) of SU(2).

\bigskip \noindent
{ Before concluding this section, we would like to make a couple of remarks regarding the computation of \emph{entanglement entropy}. Given a Hilbert space $\cH_\Gamma$ of wave functionals which factorizes into the tensor product $\cH_A \otimes \cH_{\bar A}$ of Hilbert spaces associated with a region $A$ and its complement, the computation of the entanglement entropy for a state $\psi \in \cH_\Gamma$ proceeds as follows: $(i)$ Compute the density matrix of the state $\psi$, $(ii)$ trace over the degrees of freedom in $\cH_{\bar A}$ so as to obtain the reduced density matrix, $(iii)$ compute the entanglement entropy between the region $A$ and its complement as the Von Neumann entropy of the reduced density matrix. However, an inherent feature of lattice gauge theories, such as the one under consideration here, is the non-locality of the degrees of freedom, which in turn prevents the factorization of the Hilbert space $\cH_\Gamma$. 

One solution to this issue is the so-called \emph{extended Hilbert space approach} \cite{Buividovich:2008gq, Donnelly:2008vx, Donnelly:2011hn, Donnelly:2014gva}. The basic principle is to embed states of the original Hilbert space of functionals everywhere gauge invariant into an \emph{extended} Hilbert space for which violations of the Gau{\ss} constraint are allowed at the interface between the region $A$ and its complement. More precisely, the interface is chosen so that it is transversal to the links of the graph $\Gamma$. At the intersection between the links and the interface, bivalent nodes are added at which the Gau{\ss} constraint is relaxed. It is then possible to perform the \emph{splitting} of the resulting extended Hilbert space into a tensor product.

In \cite{Donnelly:2016auv} it was shown that for any gauge theories the extended Hilbert space forms carries a representation of a boundary symmetry algebra and this boundary symmetry algebra can be understood as being dual to the projection of the constraints on the boundary. In  \cite{Delcamp:2016eya} this procedure was exemplified  for Hilbert spaces $\cH_\Gamma$ of wave functionals satisfying a set of constraints $\{\mathscr{ C}\}$, which may include or not the Gau{\ss} constraint. The splitting was then defined as dual to the gluing procedure. We saw above an example of such procedure where the gluing of two half-links required the implementation of the Gau{\ss} constrain at the bivalent node along which the gluing is performed. This lead to the definition of the fusion tensor product which  is necessary in order to create singlet states when subregions are glued together. Conversely, we can define the splitting of a link by introducing a bivalent node at which the Gau{\ss} constraint may be relaxed which would turn the corresponding fusion tensor product $\boxtimes_{\rm SU(2)}$ into the regular tensor product $\otimes$. The states resulting from this splitting are such that we recover the original one when applying the gluing procedure. In other words, in order to define the extended Hilbert space, we need to relax the constraints which would need to be enforced when performing the dual gluing procedure. This general procedure will also apply later when we will trade the Gau{\ss} constraint for the flatness constraint.

For the sake of clarity, let us present the explicit computation in a simple case. Since we have defined the kinematical space for a single link only, we will choose the graph $\Gamma = \ell $  so that region $A$ corresponds to the left half-link $\ell_L$. The procedure generalizes straightforwardly to a more general graph. First we need to define the map $\mathcal{E}$ which embeds the states $\cH_\ell$ into the corresponding extended Hilbert space such that
\begin{equation}
	\mathcal{E}: \cH_\ell \simeq \cH_{\ell_L} \boxtimes_{\rm SU(2)} \cH_{\ell_R} \longrightarrow \cH_\ell^{\rm ext} \simeq \cH_{\ell_L} \otimes \cH_{\ell_R} \; .
\end{equation}
We consider an element $|\psi \ra = |j,m,n \ra$ of an orthonormal basis of $\cH_\ell$. The corresponding embedded state explicitly reads
\begin{equation}
	\mathcal{E}(|j,m,n\ra) = \frac{1}{\sqrt{d_j}}\sum_p | j,p,n \ra \otimes | j,m,p \ra \; .
\end{equation}
The reduced density matrix of the state $| \psi \ra$ is then obtained as
\begin{equation}
	\mathcal{D}_L^\psi = {\rm tr}_{\ell_R} \big(\mathcal{E}(\psi) \overline{\mathcal{E}(\psi)} \big)
	 = \sum_{p} \frac{|j,p,n \ra \la j,p,n|}{d_j} \; .
\end{equation}
The entanglement entropy of the state $|\psi \ra$ finally reads
\begin{align}
	\nn
	S^\psi_L \equiv S_L\big(\mathcal{E}(\psi) \big) &= -{\rm tr}_L(\mathcal{D}^\psi_L {\rm log}\, \mathcal{D}^\psi_L) \\
	&=   {\rm log}\,d_j \; .
	\label{entEnt}
\end{align}
Most importantly, obtaining a non-vanishing entanglement entropy is characteristic of the fact that the symmmetry whose constraint we relaxed is described by a non-abelian group. It would indeed vanish in the case of an abelian group since the irreducible representations are all one-dimensional.}

\bigskip \noindent
So far we have defined the kinematical Hilbert space associated to a single link. The next step consists in defining the kinematical Hilbert space for a general graph $\Gamma = (\ell_1, \dots, \ell_L)$. As mentioned before, we choose this graph to be the dual of some triangulation $\triangle$ so that it only contains three-valent nodes. By assigning a state $|j,n \ra \la j,m |$ to every link $\ell \subset \Gamma$, we obain a basis for the Hilbert space $\cH^{\rm kin}_\Gamma$. The next step consists in gluing the link states together along nodes $n \subset \Gamma$ so as to obtain the \emph{spin network basis}. 

\subsection{Spin network basis}

Since we are dealing with a graph $\Gamma$ which only contains three-valent nodes, it is sufficient to define the state associated with a single three-valent node. Such state can be obtained as the gluing of three half-link states as depicted below  
\be
	\includegraphics[scale=1, valign=c]{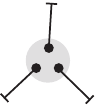}
\ee
where we follow the conventions of fig.~\ref{glueing}. As before, this gluing is performed so as to implement the Gau{\ss} constraint at the corresponding node.
We will now explain how to implement the Gau{\ss} constraint at such nodes of $\Gamma$. Let $n$ be a three-valent node, we denote three outgoing states meeting at this node by $|j_i,m_i \ra$, $i=1,\dots,3$. The vector space associated with the node is therefore given by the tensor product $V_{j_1} \otimes \cdots \otimes V_{j_3}$. In order to enforce the gauge invariance, we are looking for an invariant linear map $\iota_n^{j_1j_2j_3}: V_{j_1} \otimes V_{j_2} \otimes V_{j_3} \rightarrow \mathbbm{C}$. Such a map exists because the group SU(2) is also a Hopf algebra which comes equipped with a comultiplication map allowing to compute tensor product of representations. Its explicit expression is provided by the so-called Wigner-$3jm$ symbols\footnote{These are related to the Clebsh-Gordan coefficients via
\be 
	\Big({}^{\; j_1 \;\;\; j_2 \;\; J}_{\, m_1 \; m_2 \, M}\Big) = (-1)^{j_2-j_1-J}\frac{1 }{\sqrt{d_J}}
	C_{m_1m_2 M^*}^{j_1j_2J}
\ee
where $|J,M^*\ra= (-1)^{J-M}|J,-M\ra$ is the conjugate state.} which satisfy
\be
	\sum_{m_1,m_2,m_3}
	\Big({}^{\; j_1 \;\;\; j_2 \;\; j_3}_{\, m_1 \; m_2 \, m_3}\Big)  
	|j_1,m_1 \ra \otimes |j_2,m_2 \ra \otimes |j_3,m_3\ra = |0,0\ra \; .
\ee
It turns out that the equation above can be interpreted as a higher-valent version of the fusion tensor product. 
In the case of a higher-valent node, the intertwiner is not unique and can be obtained as a contraction of several Wigner-$3jm$ symbols. We can now define spin network states which form a basis for the Hilbert space of functionals satisfying the Gau{\ss} constraint at every node. A spin network is a triplet $(\Gamma,\{j_\ell\},\{\iota_n \})$ where
\begin{enumerate}[itemsep=0.4em,parsep=0pt,leftmargin=2em]
	\item[$\circ$] $\Gamma$ is a oriented graph embedded on $\Sigma$ . 
	\item[$\circ$] $\{j_\ell\}$ is a set of group variables labeling the links $\ell$ of $\Gamma$ .
	\item[$\circ$] $\{\iota_n\}$ is a set of intertwiners labeling the nodes $n$ of $\Gamma$ living in the invariant subspace of the tensor product of the representations spaces associated with the incoming and outgoing edges attached to $n$ {\it i.e.}
	\be
		\iota_n: \bigotimes_{\ell:n=t(\ell)}V_{j_\ell} \longrightarrow 
		\bigotimes_{\ell:n=s(\ell)}V_{j_\ell} \; .
	\ee
\end{enumerate}
A spin-network state is finally obtained by contracting the spin-$j$ states living on the links with the chosen intertwiners, the contraction pattern being dictated by the choice of graph:
\be
\Psi[\Gamma, \{j_\ell\}, \{\iota_n\}] \equiv {\rm tr}_{\{V_j\}} \Big[\bigotimes_\ell |j_\ell \rangle \langle j_\ell | \otimes \bigotimes_n \iota_n \Big] \;.
\ee 
States satisfying the Gau{\ss} constraint can then be obtained as a superposition of spin network states and the corresponding Hilbert space is denoted $\mathcal{H}_\Gamma^\mathcal{G}$.

~\\
{\bf {\small Graphical calculus:}}\\
Before pursuing with our construction, let us introduce a graphical calculus\footnote{The graphical calculus presented here is extensively usd in the condensed matter litterature in the context of Levin-Wen models \cite{Levin:2004mi}. It turns out that the spin network basis corresponds to a so-called string net model for the Rep($G$) fusion category.} for spin network states ({\it cf} for instance \cite{Levin:2004mi, DGTQFT}). As we explained in detail earlier, to each link of the graph $\Gamma$, we assign a state $|j,m,n\ra$ so that we have the following correspondence

\be
	|j,m,n \ra \equiv |j,n \ra \la j,m| \equiv \includegraphics[scale=1,valign=c]{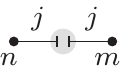} 
	 \; ,
\ee
where we implicitly make use of the self-duality of the irreducible representations of the group in order to work with non-oriented links. Furthermore, we introduce a notation for the evaluated link, i.e. so that the corresponding holonomy is trivial:
\be
	(\mathbbm{1} | j,m,n > \equiv \includegraphics[scale=1,valign=c]{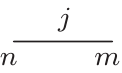} \equiv
	\includegraphics[scale=1,valign=c,raise=0.35em]{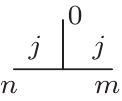} \; .
\ee
For a closed spin network, we define its \emph{evaluation} as the number obtained by contracting the intertwiners together which effectively amounts to set all the holonomies to the identity. 
We introduce the so-called \emph{$F$-symbols} 
\begin{eqnarray}\label{6j}
	F^{j_1j_2j_5}_{j_3j_4j_6} &:=& (-1)^{j_1+j_2+j_3+j_4}\sqrt{d_{j_5}d_{j_6}}
	\Big\{{}^{\, j_1 \; j_2 \; j_5}_{\, j_3 \; j_4 \,j_6}\Big\} 
\end{eqnarray}
which are defined in terms of the Wigner-$6j$ symbols 
whose expressions in terms of the $3jm$-symbols are given by
\begin{equation}\nn
	\Big\{{}^{\, j_1 \; j_2 \; j_5}_{\, j_3 \; j_4 \,j_6}\Big\}
	=  \hspace{-0.5em}\sum_{m_1,\dots,m_6} \hspace{-0.5em}
	(-1)^{\sum_{k = 1}^6 (j_k - m_k)}
	\Big({}^{\; j_1 \;\;\; j_2 \;\;\;\; j_5}_{\, m_1 \; m_2 \, -m_5}\Big) 
	\Big({}^{\;\;\; j_1 \;\;\;\, j_4 \;\;\; j_6}_{\, -m_1 \; m_4 \, m_6}\Big)
	\Big({}^{\; j_3 \;\;\;\;\, j_4 \;\;\, j_5}_{\, m_3 \; -m_4 \, m_5}\Big) 
	\Big({}^{\;\;\; j_3 \;\;\;\;\;\; j_2 \;\;\;\;\; j_6}_{\, -m_3 \; -m_2 \, -m_6}\Big) \; .   
\end{equation}
In the absence of curvature, spin network states become trivial and reduce to the evaluation of the corresponding spin network. Such evaluated spin networks are invariant under a set of local unitary transformations. The first transformation is the analogue of the 2-2 Pachner move. More precisely, the $F$-symbols perform the change of basis between ${\rm Hom}(j_1 \otimes j_4, j_2 \otimes j_3)$ and ${\rm Hom}(j_1 \otimes j_2 , j_3 \otimes j_4)$ according to the formula
\be
	\includegraphics[scale=1,valign=c]{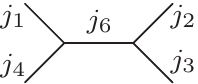}
	 = 
	\sum_{j_5}F^{j_1j_2j_5}_{j_3j_4j_6}\;
	\includegraphics[scale=1,valign=c]{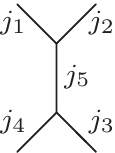} \; .
\ee
We will refer to such operation as an \emph{$F$-move}.
From the definition of the Wigner-$6j$ symbols we know that $F^{j_1j_2j_5}_{j_2j_1\,0} = \frac{v_{j_5}}{v_{j_1}v_{j_2}}$ where $v_j^2 := (-1)^{2j} {d_j}$. Using this result, we deduce the following identity as a special case of the $F$-move:
\be
	\includegraphics[scale=1,valign=c]{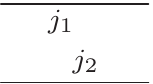}
	= 
	\sum_{j_3}\frac{v_{j_3}}{v_{j_1}v_{j_2}}\;
	\includegraphics[scale=1,valign=c]{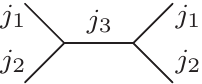} \; .
\ee
More generally, the following property holds
\be
	\label{jj}
	\widehat{h}^{j_2} \triangleright 
	\includegraphics[scale=1,valign=c,raise=0.5em]{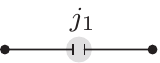} \equiv
	\includegraphics[scale=1,valign=c]{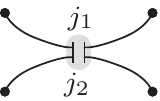}
	= \sum_{j_3}\frac{v_{j_3}}{v_{j_1}v_{j_2}}\;
	\includegraphics[scale=1,valign=c]{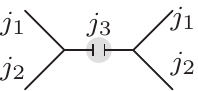} \; .
\ee
where $\widehat{h}^{j_2}$ is the holonomy operator in the spin-$j_2$ representation.
Again in the absence of curvature, we can evaluate a closed loop in order to perfom the following \emph{bubble move}
\be
	\includegraphics[scale=1,valign=c]{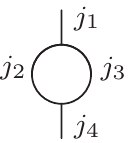} = 
	\frac{v_{j_2}v_{j_3}}{v_{j_1}}N_{j_1j_2j_3} \delta_{j_1,j_4} \;
	\includegraphics[scale=1,valign=c]{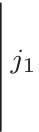}
\ee
where $N_{j_1j_2j_3}$ are the fusion rules of the fusion category ${\rm Rep}(G)$ dictating the recoupling of two irreducible representations such that $V_{j_2} \otimes V_{j_3} = \oplus_{j_1}N_{j_1j_2j_3}V_{j_1}$.\footnote{Remember that as the level of the states, this recoupling is performed by the Clebsch-Gordan coefficients, from which we define the $3jm$-symbols.} Finally, we will make extensive use of the following evaluation
\be
	\label{combo}
	\includegraphics[scale=1,valign=c]{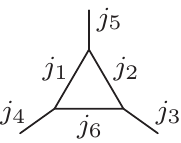}=
	\frac{v_{j_1}v_{j_2}}{v_{j_5}}F^{j_1j_2j_5}_{j_3j_4j_6}\;
	\includegraphics[scale=1,valign=c]{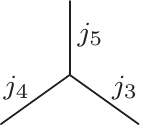} 
\ee
which is simply a combination of an $F$-move and a bubble move.

\subsection{Dynamics}
Let us now turn to the implementation of the dynamics. This means characterizing the space of states solving the flatness constraint and defining the physical inner product. This can be done by turning the constraint into a projector. This is particularly easy when dealing with wave functionals $\psi(g)\equiv (g|\psi\ra $ as it suffices to set all the plaquette holonomies to the identity. More precisely, for a state $|\psi_\Gamma \ra \in \mathcal{H}_\Gamma^\mathcal{G}$, we have
\begin{align}
	\label{projflat}
	{\mathbbm{P}}^{\mathcal{F}} \; : \; \mathcal{H}_\Gamma^\mathcal{G} &\longrightarrow \mathcal{H}_\Gamma^{\rm phys} \\ \nn
	| \psi_\Gamma \ra &\longmapsto \prod_{p \subset \Gamma}
	{\mathbbm{P}}^{\mathcal{F}}_p|\psi_\Gamma \ra
\end{align}
with ${\mathbbm{P}}^{\mathcal{F}}_p| \psi_p \ra = |\delta(g_p,\mathbbm{1})\psi_p \ra$ and the inner product is formally provided via
\be
	\big\langle \, {\mathbbm{P}}^{\mathcal F} \psi_2 \; | \
	{\mathbbm{P}}^{\mathcal F} \psi_1 \, \big\rangle_{\rm phys} = 
	\big\langle \, \psi_2 \; | \;
	{\mathbbm{P}}^{\mathcal F} \psi_1 \, \big\rangle_{\rm kin} \; .
\ee 
 We would like to perform such a projection directly in the spin network basis, but starting from holonomy states \eqref{genFT}. Let us consider a string $\mathfrak{s}$ of consecutive links labeled by the group variables $\{g_\ell\}$. We define an operator $\mathbbm{P}^j$ labeled by the spin-$j$ representation of $G$ acting on $\mathfrak{s}$ as follows
\be
	\mathbbm{P}^j|\{g_\ell\}) := {\rm tr}_{\{V_j\}}\Big[\prod_{\ell \subset \mathfrak{s}} D^j(g_\ell)\Big]| \{g_\ell\} ) = 
	\chi^j\Big(\prod_{\ell \subset \mathfrak{s}}g_\ell\Big)| \{g_{\ell}\} ) \; .	
\ee
Using that the tensor product of irreducible representations decomposes as
\be
	D^{j_1}(g) \otimes D^{j_2}(g) = \bigoplus_{j_3} N_{j_1j_2j_3} D^{j_3}(g) \;,
\ee
we deduce the operator product of two operators 
\be
	\mathbbm{P}^{j_1}\mathbbm{P}^{j_2}= \sum_{j_3}N_{j_1j_2j_3}\mathbbm{P}^{j_3} \; .
\ee
Using the fact that the group delta function is provided by
\be
	\delta(g) = \sum_{j}d_j \chi^j(g)\;,
\ee
it is clear that the operator $\mathbbm{P}:=\sum_j d_j \mathbbm{P}^j$ defines a projection onto the subspace of states satisfying $\prod_{\ell \subset \mathfrak{s}}g_{\ell} = \mathbbm{1}$. It is however not a ``full'' projector, {\it i.e.}, in the sense that the square would give again the projector, since
\be
	\mathbbm{P}^2 =\sum_{j_1,j_2}d_{j_1}d_{j_2}\mathbbm{P}^{j_1}\mathbbm{P}^{j_2} = 
	\sum_{j_1,j_2,j_3}d_{j_1}d_{j_2}N_{j_1j_2j_3}\mathbbm{P}^{j_3}= \Big(\sum_{j_2} d_{j_2}^2 \Big)\sum_{j_3}d_{j_3} \mathbbm{P}^{j_3} = \delta(\mathbbm{1})\mathbbm{P}
\ee
where we used that $\sum_{j_1}N_{j_1j_2j_3}d_{j_1} = d_{j_2}d_{j_3}$. As we will see, this is reminescent of the divergences appearing in the Ponzano-Regge model \cite{Ponzano_Regge_1969, Barrett:2008wh, Freidel:2004vi, Livine:2016vhl}. 

We now would like to consider the case of a closed string in order to impose the flatness constraint around a plaquette. We denote the corresponding operator $\mathbbm{P}_p$. Its action in the spin network basis on a triangular plaquette can be computed using the graphical calculus as follows
\begin{align} \nn
	\mathbbm{P}_p^J \triangleright \hspace{-1em}
	\includegraphics[scale=1,valign=c]{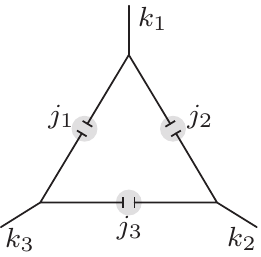} &= 	
	\includegraphics[scale=1,valign=c]{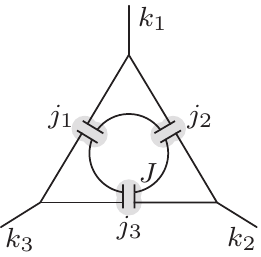}
	=  \sum_{l_1,l_2,l_3}\frac{v_{l_1}}{v_{j_1}v_J}
	\frac{v_{l_2}}{v_{j_2}v_J}\frac{v_{l_3}}{v_{j_3}v_J}
	\includegraphics[scale=1,valign=c]{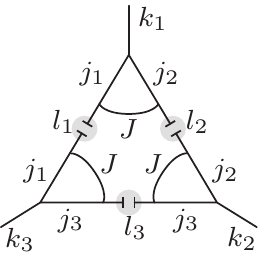} \\
	&=  \sum_{l_1,l_2,l_3}F^{j_1j_2k_1}_{l_2l_1J}F^{j_2j_3k_2}_{l_3l_2J}
	F^{j_3j_1k_3}_{l_1l_3J}\frac{v_{l_1}v_{l_2}v_{l_3}v_{j_1}v_{j_2}v_{j_3}}{v_{k_1}v_{k_2}v_{k_3}v_J^3}
	\includegraphics[scale=1,valign=c]{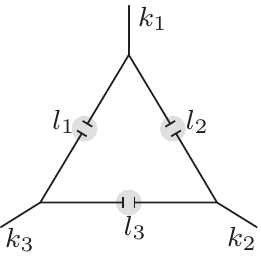} 
\end{align} 
where we repeatedly made use of the properties \eqref{jj} and \eqref{combo}. It is possible to use such properties since the curvature is contained within the loop labeled by $J$, outside of it, we can perform an evaluation of the spin network and use the local unitary transformations.
Using the definition for the $F$-symbols and the factors $v$, the action of $\mathbbm{P}_p$ can be rewritten
\be \nn
	 \sum_{l_1,l_2,l_3}(-1)^{J+\sum_{i=1}^3j_i+k_i+l_i}
	\bigg(\prod_{i=1}^3\sqrt{d_{l_i}d_{j_i}}\bigg)
	\Big\{{}^{\, j_1 \; j_2 \; k_1}_{\, l_2 \;\, l_1 \; J}\Big\}
	\Big\{{}^{\, j_2 \; j_3 \; k_2}_{\, l_3 \;\, l_2 \; J}\Big\}
	\Big\{{}^{\, j_3 \; j_1 \; k_3}_{\, l_1 \;\, l_3 \; J}\Big\}
	\includegraphics[scale=1,valign=c]{fig/opertri4S-eps-converted-to.pdf} \;.
\ee

~\\
{\bf {\small Ponzano-Regge model:}}\\
We now wish to relate the action of such an operator with the Ponzano-Regge model whose definition we will now recall. Let $\mathcal{M}$ be a compact three-dimensional manifold and $\partial \mathcal{M}$ its boundary. We consider a triangulation of the boundary denoted $\partial \large\triangle$ and $\large\triangle$ a triangulation of $\mathcal{M}$ which is compatible with $\partial \large\triangle$. The triangulation $\large\triangle$ is made of 3-simplices $\sigma_3$, triangles $\sigma_2$, edges $\sigma_1$,
and vertices $\sigma_0$. Let $\{j_{\large\triangle}\}$ be an admissible spin coloring of the edges of $\large\triangle$, and $\{j_{\partial \large\triangle}\}$ an admissible coloring of the edges of the boundary triangulation which agrees with $\{j_{\large\triangle}\}$. We can now define the following state sum
\begin{equation}
	\label{PR}
	\mathcal{Z}_{\rm PR}(\mathcal{M},\large\triangle,\{j_{\partial \large\triangle}\}) = \sum_{\{j_{\large\triangle}\}} \prod_{\sigma_1}(-1)^{2j}d_j\prod_{\sigma_3}(-1)^{\sum_{i=1}^6j_i}
	\Big\{{}^{\, j_1 \; j_2 \; j_5}_{\, j_3 \; j_4 \, j_6}\Big\} \; .
\end{equation}
The path integral $\mathcal{Z}_{\rm PR}$ acts as a projector on a spin network state by performing its evaluation \cite{Ooguri:1991ni, Rovelli:1993kc}. More precisely, in the case where the manifold $\mathcal{M}$ has boundaries, if we think of the components of $\partial \large\triangle$ as dual to graphs $\Gamma$, then $\mathcal{Z}_{\rm PR}$ computes the transition amplitude between the corresponding spin network states. For instance, in the case of the triangular plaquette, one has
\begin{align}
	\Bigg(\mathcal{Z}_{\rm PR} \triangleright
	\includegraphics[scale=1,valign=c]{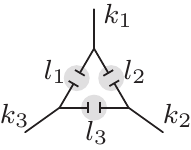}\Bigg)
	&:=F^{l_1l_2k_1}_{k_2k_3l_3}\frac{v_{l_1}v_{l_2}}{v_{k_1}}
	\Bigg(\mathcal{Z}_{\rm PR} \triangleright \includegraphics[scale=1,valign=c]{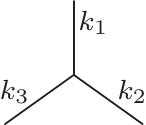} \Bigg) \\
	& = 
	(-1)^{k_1+k_2+k_3}\sqrt{d_{l_1}d_{l_2}d_{l_3}}
	\Big\{{}^{\, l_1 \; l_2 \; k_1}_{\, k_2 \; k_3 \, l_3}\Big\}
	\Bigg( \mathcal{Z}_{\rm PR} \triangleright \includegraphics[scale=1,valign=c]{fig/bubbleFmovePR3S-eps-converted-to.pdf} \Bigg) \;
\end{align}
Using the following well-know identity ({\it cf} for instance \cite{Livine:2016vhl}) which corresponds to the 4-1 Pachner move
\be\label{4-1}
	d_J 
	\Big\{{}^{\, j_1 \; j_2 \; k_1}_{\, k_2 \; k_3 \, j_3}\Big\}
	= \sum_{l_1,l_2,l_3}(-1)^{J+\sum_{i=1}^3j_i+k_i+l_i}d_{l_1}d_{l_2}d_{l_3}
	\Big\{{}^{\, j_1 \; j_2 \; k_1}_{\, l_2 \;\, l_1 \; J}\Big\}
	\Big\{{}^{\, j_2 \; j_3 \; k_2}_{\, l_3 \;\, l_2 \; J}\Big\}
	\Big\{{}^{\, j_3 \; j_1 \; k_3}_{\, l_1 \;\, l_3 \; J}\Big\}
	\Big\{{}^{\, l_1 \; l_2 \; k_1}_{\, k_2 \; k_3 \, l_3}\Big\}				
	 \; ,	
\ee
we find that
\be
	(\mathcal{Z}_{\rm PR}\circ \mathbbm{P}^J_p)	\triangleright
	\includegraphics[scale=1,valign=c]{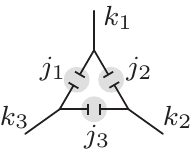} = 
	d_J\mathcal{Z}_{\rm PR}\triangleright
	\includegraphics[scale=1,valign=c]{fig/bubbleFmovePR2S-eps-converted-to.pdf} \; .
\ee
The property $ \mathbbm{P}_p\circ \mathbbm{P}^J_p = d_J  \mathbbm{P}_p$ is the property satisfied by the projection operator. 
This confirms that the Ponzano-Regge operator realises dynamically  the plaquette projection operator $\mathbbm{P}_p$. 
We see here that in the path integral picture the dichotomy is clear, the Gau\ss \,constraint is encoded in the kinematical basis of states and the curvature constraint in the choice of transition amplitude.

\section{Dual canonical quantization: $\mathcal{F} \rightarrow \mathcal{G}$} \label{F-G}
In this section, we present an alternative quantization procedure for Loop Quantum Gravity. As opposed to the usual scheme, the flatness constraint will be imposed at the kinematical level and the Gau{\ss} constraint will encode the dynamics of the theory. As we will see, this goes together with a change of basis for the kinematical Hilbert space.
The reason behind the existence of two different basis relies upon the fact that when representing the holonomy-flux algebra, we can chose to diagonalize either 
$\widehat{X} \cdot \widehat{X} $ and $\widehat{X}^3 $, which leads to the spin network basis described earlier, or we can choose to diagonalize $\widehat{h}$ which leads to the group network basis that we will now describe.

\subsection{Kinematical space$^\star$}

Let $\large\hexagon$ be a polytope decomposition of $\Sigma$ and $\large\hexagon_1=(e_1,\dots,e_E)$ its one-skeleton. To keep the exposition simple, we will assume that $\large\hexagon_1$ has only three-valent vertices. We denote the corresponding Hilbert space $\mathcal{H}_{\hexagon_1}^{\rm kin}$.	
The Poisson brackets of the holonomy-flux algebra \eqref{Poisson e} associated to an edge $e \subset \large\hexagon_1$ are turned into the following commutators which form the \emph{edge algebra} $\mathcal{A}_e$:
\be\label{loopalgbis}
	[\widehat{h}_e,\widehat{h}_e] = 0 \q , \q 
	[\widehat{X}^a_e ,\widehat{h}_e] = i \widehat{h}_e \tau^a \q , \q  
	[\widehat{X}^a_e , \widehat{X}^b_e ] = i  \epsilon^{ab}{}_{c}\widehat{X}^c_e \; .
\ee
Hence the flux $\widehat{X}_e $ acts as left invariant derivative on functionals of the holonomy $h_e$. The right invariant derivative is naturally provided by   $\widehat X_{e^{-1}}:= -h_e \widehat X_{e} h_e^{-1}$, which is the parallel transport of $\widehat{X}_e $ along the edge $e$. Similarly this operator algebra also contains $\widehat h_{e^{-1}}$.

In the LQG case, we worked in the spin representation for which the Gau{\ss} constraint has a natural action. That is we constructed the representation of the link algebra from the representations of the flux sub-algebra. In order to have a natural implementation of  the flatness constraint, we are going to consider a representation of the edge algebra built from the representation of the other natural sub-algebra, i.e. the \emph{holonomy algebra}, spanned by the matrix element operators of $\widehat h_e$ and $\widehat h_{e^{-1}}$. This algebra is commutative\footnote{Switching on the cosmological constant will change this and would therefore provide more interesting structures. } so its irreducible representations are one-dimensional. A given representation ${V}_h$ is parametrized by the holonomy $h\in {\rm SU(2)}$ and diagonalizes the holonomy operator 
\be
	\hat{h} |h\ra = {h}|h\ra \; .
\ee
Note that these states are not normalizable since the scalar product is distributional, i.e.
\be
	\la h| g\ra = \delta_{h^{-1} g}
\ee
where $\delta_{{\sss \bullet}}$ denotes Dirac delta function. 
This is a manifestation that we are dealing with a non-compact group, namely the (deformed) group of translations. 

\medskip \noindent
Previously, as part of the spin network construction, we defined link states by identifying bimodules states $|j,m,n \ra$, which carry both a right and a left group action, with operators $|j,n \ra \la j,m|$ such that we can think of the state $|j,n \ra$ (resp. $\la j,m |$) as being associated with the source (resp. target) endpoint and carries a left (resp. right) group action. In order to mimic this construction, we make use of a notation which distinguishes the bimodule state $|g)$ \eqref{genFT} entering the definition of the wave functional $\psi(g) \equiv (g | \psi \ra$ from the half-edge states $|g \ra$ and $\la g |$. The identification between the bimodule state and the algebra operator reads $|g ) \equiv | g \ra \la g |$.

Since the irreducible representations of the group of translations are all one-dimensional, the bimodule structure associated with an edge is particularly simple. To a given oriented edge $e$, we assign the bimodule state $|g \ra \la g | $ where the dual state is naturally associated with the target endpoint so that the operator $\widehat{h}_e$ acts from the left on $|g \ra$ and $\widehat{h}_{e^{-1}}$ acts from the right on $\la g |$. The kinematical Hilbert space associated with the edge $e$ is finally provided by
\be
	\cH_e \equiv  \int^\oplus_G {\rm d}g \, {V}_{g} \ot V_{g^{-1}} \; .
\ee
By construction the operators $\widehat{h}_e$  and $\widehat{h}_{e^{-1}}$ act diagonally on the bimodule states living in $\cH_e$: 
\be
	\widehat{h}_e |h_e\ra\la h_e| = {h}_e |h_e\ra\la h_e| \q , \q \widehat{h}_{e^{-1}} |h_e\ra\la h_e|= |h_e\ra\la h_e| h^{-1}_e \; .
\ee
The action of the fluxes cannot be written down as easily.\footnote{Similarly to the action of the momentum operator $\widehat{p}$ on the state $|x\ra$ in standard quantum mechanics.} However the action of the exponentiated version of the fluxes is easier to define since they act as a left and right translation of the group label, i.e.
\be
	\label{translation g}
	e^{i\theta \cdot \widehat{X}_e} |h_e\ra\la h_e| 
	= | h_e e^{\theta \cdot \tau}\ra\la  h_e e^{\theta \cdot \tau}| \q ,  \q  
	e^{i\theta' \cdot \widehat{X}_{e^{-1}}} |h_e\ra\la h_e| = | e^{-\theta' \cdot \tau} h_e\ra\la  e^{-\theta' \cdot \tau} h_e|
\ee
where $\tau^a$ are the $\su(2)$ generators.\footnote{This action could also be obtained using the graphical calculus (see \eqref{actionX})} One can now verify that the canonical commutation relation is satisfied. Indeed, one has 
\be
	\big(e^{i\theta \cdot \widehat{X}_e} \widehat{h}_e e^{-i\theta \cdot \widehat{X}_e} \big) |h_e\ra 
	= \big( e^{i\theta \cdot \widehat{X}_e} \widehat{h}_e \big)  |h_e e^{-\theta \cdot \tau}\ra
	=
	 h_e e^{-\theta \cdot \tau} |h_e \ra \; .
\ee
Finally, the group states can be related to the spin network states via the formula
\be
	|j,m,n\ra= \int_G {\rd} g  |g\ra D^j_{mn}(g)\la g| 
\ee
which can be interpreted as the inverse of the group Fourier transform \eqref{genFT}.

\subsection{Fusion tensor product$^\star$}

As in the LQG case, it is possible to construct the edge state $|g)$ by considering the fusion tensor product of two half-edge states. This fusion product will make apparent the fact that we are using the flatness constraint in order to define the gluing of half-edges, as opposed to the previous scenario where we used the Gau{\ss} constraint. 

We start with two half-edges referred to as $e_L$ and $e_R$ so that the corresponding half-edge algebras are denoted by $\cA_{e_L,e_R}$ and the corresponding half-edge Hilbert spaces by $\cH_{e_L,e_R}$, respectively. States living in such Hilbert spaces are denoted by
\begin{align*}
	|g_L\ra\la g_L| &\equiv 
	\includegraphics[scale=1, valign=c, raise=0.45em]{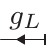} \in \cH_{e_L} \\ |g_R\ra\la g_R| &\equiv \includegraphics[scale=1, valign=c,raise=0.55em]{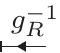}  \in \cH_{e_R} \; .
\end{align*}
where it is understood that since half-edges have two endpoints, these Hilbert spaces can be identified with bimodules. We can now implement the fusion tensor product so as to recover the Hilbert space $\cH_e$ for the full edge, i.e.
\be
	\cH_e \simeq \cH_{e_L} \boxtimes_{F[\rm{SU}(2)]} \cH_{e_R} \; ,
\ee
where the fusion is over the algebra $F[{\rm SU(2)}]$ of functions on $\rm{SU}(2)$.
In practice, we need to look at equivalence classes of states for which the action of the holonomy operator acting on the right of the left half-edge and the one acting on the left of the right half-edge are the same. In other words, the right translation for the left half-edge must match the left translation for the right half-edge. This is the \emph{matching constraint} for the dual LQG. The states
 \be
	|g_L\ra\la g_L|   \boxtimes_{F[\rm{SU}(2)]} |g_R\ra\la g_R|   
\ee 
are therefore required to satisfy
\be
	\label{equiv2}
	\big(\widehat{g}_{e_L^{-1}}|g_L\ra\la g_L|  \big)
	\boxtimes_{F[\rm{SU}(2)]} |g_R\ra\la g_R|  =  |g_L\ra\la g_L|  \boxtimes_{F[\rm{SU}(2)]} \big( \widehat{g}_{e_R} |g_R\ra\la g_R| \big) 
\ee
such that
\begin{align}
	\widehat{g}_{e_R} |g_R\ra\la g_R| &\equiv g_R |  g_R\ra\la g_R| \\
	\widehat{g}_{e_L^{-1}}|g_L\ra\la g_L|   &\equiv |g_L\ra\la g_L| g^{-1}_L \;  .
\end{align}
 Equivalence \eqref{equiv2} implies that the fusion tensor product projects down to states that have matching $g_L$ and $g_R$, and on zero otherwise. This is precisely the definition of a bivalent intertwiner which implements the flatness constraint at the bivalent vertex along which the gluing of the half-edges is performed, i.e. 
\begin{align}
	|g_L\ra\la g_L|   \boxtimes_{F[\rm SU(2)]} |g_R\ra\la g_R|   &\sim  |g_L\ra\la g_L| g_R\ra\la g_R|  \\ &=  |g_L\ra  \la g_R| \, \delta_{g_L^{-1}g_R} \; .
\end{align}
Therefore we recover the states for the full edge which we can now depict as
\be
	|g \ra \la g | \equiv 	
	\includegraphics[scale=1, valign=c, raise=0.5em]{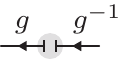}
\ee
where the gray dot represents the matching constraint.  

\bigskip \noindent
{As before, the gluing procedure we just introduced can be dualized so as to define a notion of splitting which in turn could be used to perform computations of entanglement entropy for group networks. The strategy is exactly the same as for spin networks. In order to split an edge we need to relax the very same constraint which needs to be implemented when performing the gluing of the half-edges, namely the flatness constraint. More in detail, considering an edge $e$ and its corresponding Hilbert space $\cH_e \simeq \cH_{e_L} \boxtimes_{F[\rm{SU}(2)]} \cH_{e_R}$, we are looking for an embedding map 
\begin{equation}
	\mathcal{E}: \cH_e \simeq \cH_{e_L} \boxtimes_{F[{\rm SU(2)}]} \cH_{e_R} \longrightarrow \cH_e^{\rm ext} \simeq \cH_{e_L} \otimes \cH_{e_R}
\end{equation}
where $\cH_e^{\rm ext}$ is the extended Hilbert space such that the flatness constraint may be violated at the bivalent vertex along which the splitting is performed. Since the group of translations is abelian, the embedding map takes a very simple form. Indeed, given a basis state $|g ) \in \cH_e$, it reads $\mathcal{E}\big(|g)\big) = |g ) \otimes |g)$. The computation of the density matrix and the entanglement entropy would then follow the same steps as before. However, because our states are now parametrized by a Lie group element, the final result would typically diverge.}

\medskip
\noindent 
So far we have defined the kinematical Hilbert space associated to a single edge. We can further define the kinematical Hilbert space for a general graph which we choose to be the one-skeleton $\large\hexagon_1 = (e_1, \dots, e_E)$ of a polytope decomposition $\large\hexagon_1$. As mentioned before, we choose this polytope decomposition so that it only contains three-valent vertices. By assigning a state $|g \ra \la g |$ to every edge $e \subset \large\hexagon_1$, we obain a basis for the Hilbert space $\cH^{\rm kin}_{\hexagon_1}$. The next step consists in gluing the edge states together along vertices $v \subset \large\hexagon_1$ so as to obtain the \emph{group network basis}.

\subsection{Group network basis}

{
Let us know implement the flatness constraint at every vertex $v \subset \large\hexagon_1$. Since the graph $\large\hexagon_1$ only contains three-valent vertices, it is enough to define the state associated with a single vertex. More general states can then be obtained as a gluing of such states as previously defined. Denoting by $|g_1\ra,|g_2\ra,|g_3 \ra$, three states meeting at a vertex $v$, the implementation of the flatness constraint should tell us that the state resulting from the gluing of these three states at $v$ vanishes if the product of the corresponding holonomies is not the identity. As before, we  introduce intertwiners to build explicitly the corresponding state which are the analogues of the Clebsch-Gordan coefficients. However, at the difference of the SU(2) spin intertwiners, we need to take special care when distinguishing a given intertwiner and its dual map. We define the map 
$\iota_v^{g_1,g_2}:  V_{g_1} \otimes V_{g_2} \rightarrow V_{g_1g_2}$ such that
\begin{equation}
	\label{interG}
	\iota_v^{g_1,g_2} : |g_1 \ra \otimes |g_2 \ra \, \longmapsto \, \beta(g_1,g_2)  |g_1g_2 \ra 
\end{equation}
where $\beta(g_1,g_2)$ is a phase factor\footnote{The phase factor $\beta(g_1,g_2)$ should be reminiscent of the fact that the Clebsh-Gordan coefficients for SU(2) are only defined up to a phase.} which we could choose to set to the identity, but it is convenient  not to for the time being. This defines the Clebsh-Gordan coefficient that interwinnes the non-cocomutative co-product: $ \iota_v^{g_1,g_2} \Delta \widehat{g}_{AB} = \widehat{g}_{AB} \iota_v^{g_1,g_2}$ where $\Delta \widehat{g}_{AB} =\sum_C (\widehat{g}_{AC}\otimes \widehat{g}_{CB})$ is the coproduct dual to the SU$(2)$ product law.

In order to construct invariants we need the dual map 
$\bar\iota^{g_1,g_2,g_3}_v : V_{g_3} \rightarrow V_{g_1} \otimes V_{g_2}$. This map  involves a distribution and is  defined according to 
\begin{equation}
	\label{interGd}
	\bar\iota_v^{g_1,g_2,g_3} : |g_3 \ra \, \longmapsto \, \frac{\delta_{g_1g_2,g_3}}{ \beta(g_1,g_2)}|g_1 \ra \otimes |g_2 \ra\; .
\end{equation}
This is the generalization of the embedding map associated with the fusion tensor product to the case of the gluing of three half-edges. It is clear from this definition that the map 
$\iota_v^{g_1,g_2}$ is  finite while the dual map $\bar\iota_v^{g_1,g_2,g_3 }$ is distributional. 

\medskip\noindent
Similarly to the spin network basis, we can define what we will refer to as the \emph{group network basis} for the Hilbert space of functionals satisfying the flatness constraint at every vertex. A group network is a triplet $(\large\hexagon_1,\{g_e\},\{\iota_v\})$  where
\begin{enumerate}[itemsep=0.4em,parsep=0pt,leftmargin=2em]
	\item[$\circ$] $\large\hexagon_1$ is the one-skeleton of a polygon decomposition embedded on $\Sigma$.
	\item[$\circ$] $\{g_e\}$ is a set of group variables labeling the edges $e$ of $\large\hexagon_1$. 
	\item[$\circ$] $\{\iota_v\}$ is a set of ``intertwiners'' labeling the vertices $v \subset \large\hexagon_1$ which impose that the oriented product of holonomies attached to a vertex $v$ must be the group identity. 
\end{enumerate}
A group network state is finally obtained by considering the tensor product of the holonomy-$g$ states living on the edges with the corresponding fusion rules
\be
	\Psi^\star[{\large\hexagon_1}, \{g_e\}, \{\iota_v\}] \equiv \bigotimes_e |g_e \ra\la g_e  | \otimes \bigotimes_v \iota_v \;.
\ee 
Kinematical states satisfying the flatness constraint can then be obtained as a superposition of group network states and the corresponding Hilbert space is denoted by $\mathcal{H}^{\hexagon_1}_\mathcal{F}$.

~\\
{\bf {\small Graphical calculus$^\star$:}}\\
The group network basis also comes with a graphical calculus\footnote{In the same way the spin network basis mimics string net models \cite{Levin:2004mi} built on the Rep($G$) fusion category of the finite group $G$, the group network basis is the analogue of string net models built on the ${\rm Vec}_G$ fusion category.} whose main identities are presented below. As explained in detail earlier, to each oriented edge of the one-skeleton $\hex_1$, we assign $ |g \ra\la g  | $ and thus, we have the following correspondences
\be
	|g ) \equiv 
	\includegraphics[scale=1, valign=b, raise=0.3em]{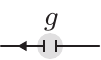}
	\equiv |g \ra \la g | \equiv 	
	\includegraphics[scale=1, valign=b, raise=0.3em]{fig/half4bis-eps-converted-to.pdf}
	\; .
\ee
We further introduce a notation for the \emph{evaluated} edge state which does not carry any torsion degrees of freedom: 
\be\label{identity}
	\includegraphics[scale=1,valign=B]{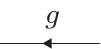} \; = \;
	\includegraphics[scale=1,valign=B]{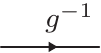} \; .
\ee
Given a vertex $v$, the number of incoming and outgoing states determines whether we have an intertwining map as defined in \eqref{interG} or its dual version \eqref{interGd}. The former case occurs when there are two incoming and one outgoing edges, and inversely for the latter case. Graphically, we distinguish these two situations as follows
\begin{equation}
	\iota_v^{g_1,g_2}=\includegraphics[scale=1,valign=c]{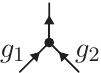} \q , \q
	\bar\iota_v^{g_1,g_2,g_3}= \includegraphics[scale=1,valign=c]{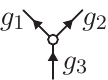} 
\end{equation}
We can express in which sense the corresponding intertwining maps are inverse of each other via the following \emph{bubble move} 
\be
	\includegraphics[scale=1,valign=c]{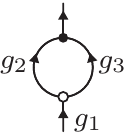} = 
	\delta_{g_1,g_3g_2} \;
	\includegraphics[scale=1,valign=c]{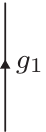} 
\ee
where we omit the labeling for one of the egdes as it follows from the definition of the black intertwiner that it is simply given by $g_3g_2$. Henceforth, we will apply the same convention for the rest of the diagrams. 
In the absence of torsion, group network states can be evaluated. As for the spin network basis, evaluated group networks are invariant under a set of local unitary transformations. The first one which is the analogue of the 2-2 Pachner move corresponds to a change of recoupling. There are several relations depending on the distribution of intertwiners. For instance, we have
\be
	\label{defAlpha}
	\includegraphics[scale=1,valign=c]{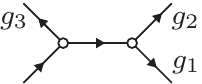}
	= 
	\; \alpha(g_1,g_2,g_3)\;
	\includegraphics[scale=1,valign=c]{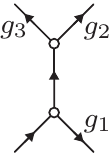} 
\ee
where $\alpha(g_1,g_2,g_3) := \beta(g_2,g_3)\beta(g_1,g_2g_3)\beta(g_1,g_2)^{-1}\beta(g_1g_2,g_3)^{-1}$, and
\be
	\label{FmoveG}
	\includegraphics[scale=1,valign=c]{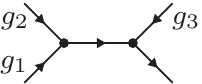}
	= 
	\; \alpha(g_1,g_2,g_3)^{-1}\;
	\includegraphics[scale=1,valign=c]{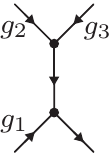} \; .
\ee
Combining this second move with the bubble move defined above, we obtain the analogue of the 3-1 Pachner move
\be
	\label{comboG}
	\includegraphics[scale=1,valign=c]{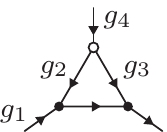}=
	\alpha(g_1,g_2,g_3)^{-1}\;
	\includegraphics[scale=1,valign=c]{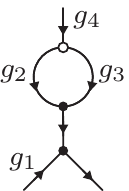} = \alpha(g_1,g_2,g_3)^{-1}\delta_{g_4,g_2g_3} \;\;
	\includegraphics[scale=1,valign=c]{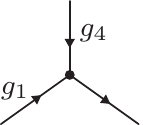} \; .
\ee
Finally, using this graphical calculus, the action of the exponentiated flux operator reads
\be
	\label{actionX}
	e^{i \theta_2 \cdot \widehat{X}} \triangleright
	\includegraphics[scale=1,valign=c,raise=0.5em]{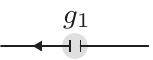} \equiv
	\includegraphics[scale=1,valign=c]{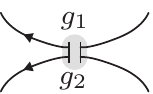}
	= 
	\int_G {\rd}g_{3}\;\; \includegraphics[scale=1,valign=c]{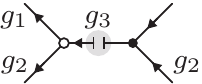} 
\ee
where $g_2 = e^{i\theta_2 \cdot \tau}$.
}

\subsection{ Dynamics$^\star$}
Let us now turn to the implementation of the dynamics. The flatness constraint having been implemented at the kinematical level, we need to find the group network states satisfying the Gau{\ss} constraint. In particular, we are looking for a face operator in the group network basis which projects onto the physical Hilbert space. The derivation will follow the same steps as before. 

Previously, in order to define the action of the operator $\mathbbm{P}_p$, we started from the action of the projector on wave functionals $\psi(g) = (g | \psi \ra$. The reason was that the flatness constraint is very simple in terms of group variables as it boils down to a group delta function. Analogously, we will see how there is a representation in which the Gau{\ss} constraint takes a similar form, namely the flux representation. Let $f$ be a triangular face of $\hex_1$. The group variables associated to the edges on the boundary of $f$ are denoted $g_1$, $g_2$ and $g_3$ such that the edges $e_1$ and $e_3$ are oriented anticlockwise and the edge $e_2$ is oriented clockwise. Denoting $|g_1,g_2,g_3 )$ the state associated to such face, we can design an operator which enforces the gauge invariance around that face via a group averaging:

\be
	\label{groupavr}
	{\mathbbm{P}}_f^{\mathcal{G}}
	|g_1,g_2,g_3 ) = \int_G {\rd} k \;
	|g_1k \, , \, k^{-1}g_2 \, , \, g_3k )  \; .
\ee
For an arbitrary face $f$ the defining formula of the operator reads
\begin{align}
	\label{groupavrGen}
	{\mathbbm{P}}_f^{\mathcal{G}} = \int_G {\rm d}k
	\bigg( \bigotimes_{e: \, {\rm or}(e)=\circlearrowleft} R_k^e\bigg) \otimes
	\bigg(\bigotimes_{e: \, {\rm or}(e)=\circlearrowright} L_k^e \bigg)
\end{align}
where the function ${\rm or}({\sss \bullet}) = \circlearrowright, \circlearrowleft$ determines the orientation of an edge with respect to the face $f$ and $R_k^e$, $L_k^e$ denotes the right and left group action, respectively.
More generally one has the following projector
\begin{align}
	{\mathbbm{P}}^{\mathcal{G}} \; : \; \mathcal{H}_{\hexagon_1}^\mathcal{F} &\longrightarrow \mathcal{H}_{\hexagon_1}^{\rm phys} \\ \nn
	| \psi_{\hexagon_1} \ra &\longmapsto \prod_{f \subset \hexagon_1}
	{\mathbbm{P}}_f^{\mathcal{G}}|\psi_{\hexagon_1} \ra \; .
\end{align}
The physical inner product is then formally provided via
\be
\big\langle \, {\mathbbm{P}}^{\mathcal G} \psi_2 \; | \
{\mathbbm{P}}^{\mathcal G} \psi_1 \, \big\rangle_{\rm phys} = 
\big\langle \, \psi_2 \; | \;
{\mathbbm{P}}^{\mathcal G} \psi_1 \, \big\rangle_{\rm kin} \; .
\ee 
We now would like to define the action of the face operator ${\mathbbm{P}}_f$, which enforces the torsion freeness around the face $f$, using the graphical calculus. To do so, it is convenient to consider wave functionals $\widetilde{\psi}(X) \equiv (X| \widetilde{\psi} \ra$ where $X \in \mathbb{R}^3$ is a flux label. In terms of such labels, the Gau{\ss} constraint associated to \eqref{groupavr} is $X_{e_1} + X_{e_2^{-1}}+ X_{e_3}=0$. In order to mimic the previous construction we would therefore need to define a delta function which enforces such Gau{\ss} constraint. This is provided by the non-commutative Fourier transform \cite{Freidel:2005ec}.

The non-commutative Fourier transform maps the space $L^2(G,{\rm d}\mu_H)$ onto the space $L^2_\star(\mathbbm{R}^3,{{\rd}}\mu)$ of functions $\mathbbm{R}^3 \sim \mathfrak{su}(2)$ equipped the Lebesgue measure and a non-commutative $\star$-product
\footnote{More exactly, as defined, it maps $L^2(\text{SO}(3),\text{d}\mu_H)$ onto a space $L_{\star}^2(\mathbbm{R}^3,\text{d}\mu)$	of functions on $\mathfrak{su}(2) \sim \mathbbm{R}^3$. One should be careful with the different coordinate patches covering $\text{SU}(2)$. Since it unnecessarily complicates the formalism, we will stick with the Fourier transform on $\text{SO}(3)$.}:
\be 
\mathsf{FT}(\psi)(X) \equiv \widetilde{\psi}(X)  \equiv (X | \widetilde{\psi} \ra := \int_G {\rd} g \, (  X|g)( g|{\psi}\ra= \int_G \text{d}g \, \text{e}_g(X) \, \psi(g) 
\ee
such that $( X | g)  = \text{e}_{g}(X)$ and $( g | X )=\overline{\text{e}_{g}(X)}$.  The plane wave is defined as
\begin{align}
	\text{e}_g: \mathfrak{su}(2) \sim \mathbb{R}^3 &\; \rightarrow \; \text{U}(1) \\
	x &\; \mapsto \; \text{e}_g(X) := e^{i\vec{p_g}\cdot \vec{X}} 
\end{align}
such that ${\rm e}_g(X)=-\frac{1}{2}{\rm tr}(|g|\vec{\tau})$. Furthermore, we can define the following non-distributional delta function 
\begin{equation}
	\delta_X(Y) = \frac{1}{8 \pi} \int {\rm d}g \, {\rm e}_{g^{-1}}(X){\rm e}_g(Y)
\end{equation}
such that $\int {\rm d}^3X \, \delta_X(Y) =1$. In particular, we have $\delta(Y):= \int {\rm d}g \, {\rm e}_g(Y)$ which is peaked on $Y=0$. We can therefore define an operator $\mathbbm{P}_f:=\int_G {\rd} k \, \mathbbm{P}^k_f$ which projects onto the subspace of states satisfying $X_{e_1} + X_{e_2^{-1}}+ X_{e_3}=0$ such that  
\be
\mathbbm{P}^k|\{X_e\}\ra := \Big[\prod_{e \subset f} {\rm e}_k({X}_e)\Big]| \{X_e\} ) = 
{\rm e}_k\Big(\sum_{e \subset f}{X}_e\Big)| \{X_e\} )	\; .
\ee
{ Its action in the group network basis on a triangular face can then be computed using the graphical calculus as follows
\begin{align} \nn
	&\mathbbm{P}^k_f \triangleright 
	\includegraphics[scale=1,valign=c]{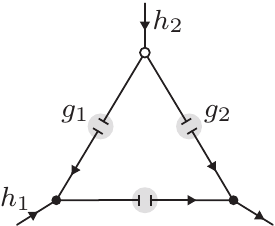} = 
	\includegraphics[scale=1,valign=c]{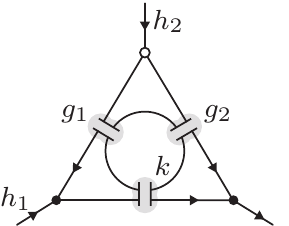}
	=  \int_{G^2}{\rd} p_1 {\rd }p_2 
	\includegraphics[scale=1,valign=c]{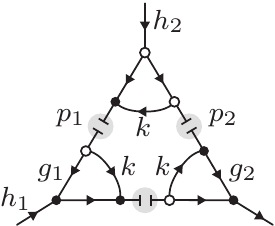} \\ \nn
	& \q = \int_{G} {\rd} p_1 {\rd}p_2 \;
	\delta_{g_1kp_1^{-1}}\delta_{g_2^{-1}kp_2} 	\;
	\alpha(h_1,g_1,k)^{-1}\alpha(h_1g_1,k,k^{-1}g_2)\alpha(g_1,k,k^{-1}g_2)^{-1}
	\includegraphics[scale=1,valign=c]{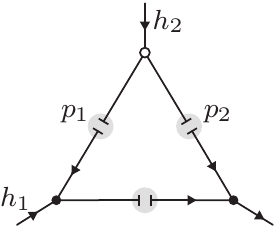} 
\end{align} 
where we repeatedly make use of equations \eqref{actionX} and \eqref{comboG}. Performing the intergrals over $G={\rm SU(2)}$, we finally obtain the following action
\begin{align} \nn
	&\mathbbm{P}^k_f \triangleright 
	\includegraphics[scale=1,valign=c]{fig/opertri1G-eps-converted-to.pdf} =
	\alpha(h_1,g_1,k)^{-1}\alpha(h_1g_1,k,k^{-1}g_2)\alpha(g_1,k,k^{-1}g_2)^{-1}
	\includegraphics[scale=1,valign=c]{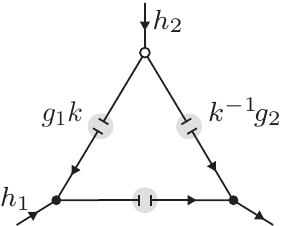} 
\end{align}
where we recognize the right and left group actions $R_k$ and $L_k$ which shift the variables $g_1$, $g_2$. Note that we recover exactly the expression \eqref{groupavr} for $\alpha \equiv 1$.
So we have obtained a graphical way to define the face operator $\mathbbm{P}_f$ which mimics the construction of the flatness operator in standard LQG. However, considering the fact that the group averaging $\eqref{groupavr}$ already allows to define such a projection in the group network basis, this approach might appear a little bit cumbersome to the reader. Nevertheless, such construction allows us to show explicitly how this construction is related to a well-know state-sum model.

~\\
{\bf {\small State sum model:}}\\
As for the standard LQG quantization scheme, we want to relate this operator to a state-sum that we will now define. Let $\mathcal{M}$ be a compact three-dimensional manifold and $\partial \mathcal{M}$ its boundary. We consider a triangulation of the boundary denoted $\partial \large\triangle$ and $\large\triangle$ a triangulation of $\mathcal{M}$ which is compatible with $\partial \large\triangle$ on the boundary. The triangulation $\large \triangle$ is made of 3-simplices $\sigma_3$, triangles $\sigma_2$, edges $\sigma_1$,
and vertices $\sigma_0$. Let $\{g_{\large\triangle}\}$ be an admissible coloring with group variables of the edges $\large\triangle$, and $\{g_{\partial \large\triangle}\}$ an admissible coloring, as prescribed by the intertwiners, of the edges of the boundary triangulation which agrees with $\{g_{\large\triangle}\}$. To a given 3-simplex $\sigma_3 = (abcd)$ such that $a < b < c < d$, we assign the topological action $\alpha(g_{ab},g_{bc},g_{cd})$, where $\alpha$ refers to the phase factor defined above. We can now define the following state sum
\begin{equation}
	\label{PRdual}
	\mathcal{Z}(\mathcal M,{\large\triangle},\{g_{\partial \large\triangle}\}) = \int \prod_{\{g_{\large\triangle}\}}{\rd}g_{\large\triangle} \prod_{\sigma_3}\alpha^{\pm1}(\sigma_3) \; .
\end{equation}
where the sign is $+1$ if the tetrahedra orientation matches the orientation of the manifold, and $-1$ otherwise.
Note that in this definition, the flatness constraint is located around the triangles as opposite to the three-valent vertices on $\hex_1$. The reason is that, as before, we define the state sum on the triangulation dual to the graph on which the basis states are defined. 
As for the Ponzano-Regge model, the path integral $\mathcal{Z}_{\rm }$ acts as a projector on a group network state by performing its evaluation. For instance, in the case of the triangular face, one has
\begin{align}
	\Bigg(\mathcal{Z}_{} \triangleright
	\includegraphics[scale=1,valign=c]{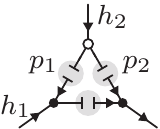}\Bigg)
	&:=\delta_{h_2,p_1p_2}\alpha(h_1,p_1,p_2)^{-1}
	\Bigg(\mathcal{Z}_{} \triangleright \includegraphics[scale=1,valign=c]{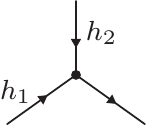} \Bigg) \; .
\end{align}
Furthermore, we have that the coefficients $\alpha$ satisfy the following relation which is the analogue of the one \eqref{4-1} verified by the Wigner-$6j$ symbols
\be	
	\label{pentagon}
	\alpha(h_1,g_1,g_2)^{-1} =\; \alpha(h_1,g_1,k)^{-1}\alpha(h_1g_1,k,k^{-1}g_2)\alpha(g_1,k,k^{-1}g_2)^{-1}\alpha(h_1,g_1k,k^{-1}g_2)^{-1} 
	\; .	
\ee
This relation directly from the definition of $\alpha$ in terms of the phase factors $\beta$.
Putting everything together,  we find that
\be\label{proj}
	(\mathcal{Z}_{}\circ \mathbbm{P}^k_f)	\triangleright
	\includegraphics[scale=1,valign=c]{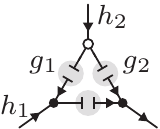} = 
	\mathcal{Z}_{}\triangleright
	\includegraphics[scale=1,valign=c]{fig/bubbleFmovePR2G-eps-converted-to.pdf} \; .
\ee
which confirms that the plaquette operator $\mathbbm{P}_f$ as defined above performs a projection in the group network basis onto the subspace of states satisfying the Gau{\ss} constraint.

\subsection{Relation with the Dijkgraaf-Witten model}\label{DW}

The alert reader may find the state-sum model that we just introduced rather familiar. It may be thought indeed as an extension for Lie groups of the Dijkgraaf-Witten state-sum model \cite{dijkgraaf1990} which is typically defined for finite groups. Let us briefly review the construction of this topological invariant as defined in \cite{dijkgraaf1990}.

Let $G$ be a finite group. We consider a topological field theory defined on the three-dimensional oriented manifold $\mathcal{M}$. We realize $\mathcal{M}$ as an union of 3-simplices $\sigma_3$ by picking a triangulation $\large\triangle$. To every 1-simplex $\sigma_1=(ab)$ of $\large\triangle$, we assign a group element $g_{ab} \in G$ such that for every 2-simplex $\sigma_2 = (abc)$, we impose the flatness condition $g_{ac} = g_{ab} \cdot g_{bc}$. The assignment of $G$-valued variables together with the flatness condition on every 2-simplex provides an admissible $G$-coloring $\{g_{\large\triangle}\}$ of $\large\triangle$. We introduce an ordering of the vertices which follows either a right-handed or a left-handed arrangement, hence it assigns an orientation for each $\sigma_3$. Because of the flatness conditions, given a 3-simplex $\sigma_3 = (a b cd)$ such that $a < b < c < d$, the $G$-coloring of $\sigma_3$ is fully determined by the coloring of the 1-simplices $(ab )$, $(b c)$ and $(cd)$. Let $\alpha : G^3 \rightarrow U(1)$ be a group 3-cocycle which represents an equivalence class in the third cohomology group $H^3(G,U(1))$. We define the topological action as $\alpha^{\pm 1}(\sigma_3) \equiv \alpha(g_{ab},g_{bc},g_{cd})^{\pm 1 }$ and the value of the exponent is either $+1$ or $-1$ whether the orientation of the 3-simplex coincides with the orientation of $\mathcal{M}$. Finally the Dijkgraaf-Witten invariant state-sum reads
\begin{equation}
	\mathcal{Z}(\mathcal M,{\large\triangle},\{g_{\partial {\large\triangle}}\}) = \frac{1}{|G|^{\#\sigma_0}}\sum_{\{g_{{\large\triangle}}\}} \prod_{\sigma_3}\alpha^{\pm 1}(\sigma_3)\; .
\end{equation}
with $\# \sigma_0$ the number of $0$-simplices in the triangulation. Apart from the finiteness of the group, this state-sum model is very similar to the one presented in the previous section which projects onto the subspace of torsion-free connections. In particular, the phase $\alpha$ appearing in \eqref{PRdual} satisfies the relation \eqref{pentagon} which is precisely the group 3-cocycle condition, as such it represents a cohomological class in $H^3(G,U(1))$. However, it turns out to be a \emph{trivial} 3-cocycle, where by trivial we mean that it belongs to the equivalence class $[\alpha^0]$ of the cocycle $\alpha^0(g_1,g_2,g_3) \equiv 1$, $\forall g_1,g_2,g_3 \in G$. Indeed, given a 3-cocycle $\alpha \in [\alpha^0]$, if $\alpha \neq \alpha^0$, then it takes the form of a 2-coboundary such that 
\be
	\alpha(g_1,g_2,g_3) \equiv d \beta(g_1,g_2,g_3) = 
	\frac{\beta(g_2,g_3)\beta(g_1,g_2g_3)}{\beta(g_1g_2,g_3)\beta(g_1,g_2)}
\ee
with $\beta$ a normalized 2-cochain. This was precisely the definition of the phase $\alpha$ in \eqref{defAlpha}. Thanks to our derivation of the operator $\mathbbm{P}_f$, we know that choosing $\alpha$ to be a trivial cocycle corresponds to enforcing the Gau{\ss} constraint. But we also know that the projector property \eqref{proj} is satisfied for any 3-cocycle $\alpha$ satisfying the identity \eqref{pentagon}. Therefore, taking a non-trivial $\alpha$ induces a twist of the Gau{\ss} constraint and we conjecture that it may correspond to a deformation of  the holomony-flux algebra into an algebra where the holonomy operators no longer commute so that the commutation of holonomies is given by a central extension.
We postpone the exploration of the role of such central extension to another paper.

}
Note finally that in the case where $G$ is finite, there is an alternative form for this state-sum which is closer related to the Ponzano-Regge model as it is expressed in terms of $6j$-symbols. More precisely, we have the following statement \cite{2006math......8614P}: Let $\mathcal{C}$ be a group category, $G$ its underlying group and $\alpha$ the associativity constraint of $\mathcal{C}$, then for all closed manifold $\mathcal{M}$
\be
	\mathsf{DW}_{G,\alpha}(\mathcal{M}) = \mathsf{TV}_\mathcal{C}(\mathcal{M})
\ee
where $\mathsf{DW}$ and $\mathsf{TV}$ stands for the Dijkgraaf-Witten and the Turaev-Viro model respectively. In condensed matter terms, this corresponds to the statement that Levin-Wen string net models for the $\text{Vec}^\alpha_G$ for given $F$-symbols can be mapped to the so-called twisted quantum double models \cite{hu2013twisted} for the corresponding 3-cocycles.

\section{Conclusion}
We presented in this paper a dual loop quantization of 3d gravity which relies upon the implementation of the flatness constraint at the kinematical level, followed by the imposition of the Gau{\ss} constraint as the dynamical constraint. The invariant states are therefore intertwiners for the (deformed) translational group. Interestingly, in order to define the kinematical Hilbert space, it was necessary to study in detail how links (resp. edges) are obtained from the gluing of half-links (resp. half-edges) together with the corresponding fusion tensor product. This study highlighted the implicit identification between states of the Hilbert space and operators of the corresponding algebra. Furthermore, we presented the corresponding state-sum model which implements the dynamics in this alternative scheme. It turns out to be related to the Dijkgraaf-Witten model which is a state-sum invariant whose input is a finite group together with a group cohomology class. Interestingly, the idea that there are two canonical quantization procedures for $BF$-theory also appears in the condensed matter litterature in the context of Levin-Wen models. Indeed, given a finite group $G$, two string-net models can be constructed, either from the category ${\rm Rep}(G)$ of finite-dimensional representations or from the category ${\rm Vec}_G$ of $G$-graded vector spaces. These two string-net models are the direct analogue of the bases we described here, namely the spin network basis and the group network basis.

Since we are dealing with intertwiners for a (deformed) non-compact group, we have to deal with regulating divergences. Note that these divergences are analoguous to the ones appearing in the Ponzano-Regge model as they arise from the imposition of the flatness constraint. This could be resolved at the kinematical level by an ad-hoc \emph{Bohr compactification}. 
However, such Bohr compactifciation scheme is not quite tenable dynamically as it would force the quantisation of the dual loop gravity model to be different from the quantization obtained by the Ponzano-Regge model. This suggests that the dynamics should resolve the quantization ambiguities. 
But this also means that we need to revise the definition of the  \emph{projective/inductive limit} of the theory \cite{DGfluxQ} since the vaccum state is non-normalisable.  We expect that this issue should be carefully studied when dealing with intertwiners for non-compact groups \cite{Freidel:2002xb, okolow}. 
Furthermore, from our experience with the Ponzano-Regge model, we know that the divergences in the partition function means that we should interpret the model as a quantum measure which allows to compute expectation values for certain observables or for certain open amplitudes. The fact that the identity does not have finite trace is not a fundamental obstruction in the definition of the model.  Finally, divergent quantities might also be the sign of some non-trivial physics, just like in the PR model \cite{Freidel2004}. The careful study of the  divergences and their meaning will be investigated elsewhere.

Changing the order of imposition of the constraints does not mean we have a new theory but merely a different perspective on the same theory. In fact, we expect that there should be a unitary map relating both models. It turns out that the gluing procedure in the spin network basis induces an algebra which is isomorphic to the group ring $\mathbb{C}[{\rm SU(2)}]$ as a vector space. Similarly, the algebra induced by the gluing in the group network basis is the abelian algebra $F[{\rm SU(2)}]$ of linear functions on SU(2). But as Hopf algebras, these two are dual to each other. Furthermore, together they form the so-called \emph{Drinfel'd double} $\mathcal{D}({\rm SU(2)}) \simeq \mathbb{C}[{\rm SU(2)}] \otimes F[{\rm SU(2)}]$ as defined in \cite{Dijkgraaf1989, 1991NuPhS..18...60D, drinfel1988}.  This was exploited in the topological order litterature to display the electric/magnetic duality of Kitaev's double model \cite{Buerschaper2010}. As a matter of fact instead of using the holonomy representation, we could represent the states differently. More specifically we can define a kinematical basis labeled by simple representations of the Drinfeld double which depends on the conjugacy classes of SU(2), and define accordingly simple intertwiners. By doing so, we would bring closer together the kinematical bases associated with each quantization scheme and make explicit the role played by the Drinfel'd double. In particular, past work \cite{Freidel:2006qv} has shown how to relate $6j$-symbols for SU(2) representations with $6j$-symbols for the simple representations of $\mathcal{D}({\rm SU(2)})$. We expect that this relation will be at the heart of the duality map. 

We noticed how the state-sum model which implements the dynamics in the dual quantization scheme resembles a DW model with a trivial 3-cocycle. In spite of the fact that the DW state-sum model requires a finite group, it suggests some interesting generalizations. Indeed, as we have already alluded, this opens the possibility  to define a generalisation corresponding to a DW model with a non trivial 3-cocycle. It would be interesting to investigate how the duality between Ponzano-Regge and DW extends to the Turaev-Viro state-sum which provides the spin foam model for Euclidean gravity with a positive cosmological model \cite{Mack:1991tg}.

All these questions are interesting but are specific to (2+1)d gravity. We hope that the present work can also give some insights to the more physically relevant (3+1)d case. From our perspective, we might think that one could  look at a different order of implementation of the constraints, that is we could implement first the (spatial) diffeomorphisms constraint and then the Gau{\ss} constraint\footnote{See also \cite{Wieland:2013cr}, where torsion appeared as a dynamical constraint in the 4d discrete picture. }. Such different implementation could be backed up by a careful analysis of the different possible discretizations, as done in the 3d case \cite{Dupuis:2017otn}. It would be interesting to see whether the intuition we developed based on teleparallel gravity could also be useful in this setting.  This is work in progress. 

\newpage

\acknowledgments
CD is supported by a NSERC
grant awarded to Bianca Dittrich. This research was supported in part by Perimeter Institute for Theoretical Physics. Research
at Perimeter Institute is supported by the Government of Canada through the Department of Innovation, Science and
Economic Development Canada and by the Province of Ontario through the Ministry of Research, Innovation and
Science.

\bibliographystyle{JHEP}
\bibliography{LQGd}

\end{document}